\documentclass[twocolumn,citeautoscript,pra,superscriptaddress,amsmath,amssymb,8pt]{revtex4-1}

\usepackage{graphicx}
\usepackage{gensymb}
\usepackage{color}
\usepackage[dvipsnames]{xcolor}
\usepackage{upgreek}
\usepackage{float}
\usepackage{upgreek}
\usepackage{amsmath,amssymb}

\newcommand{\ket}[1]{\ensuremath{\left| #1 \right\rangle}}

\begin{document}

\title{Comparison of different methods of nitrogen-vacancy layer formation in diamond for widefield quantum microscopy}

\author{A. J. Healey}
%\email{ahealey1@student.unimelb.edu.au}
\affiliation{School of Physics, University of Melbourne, Parkville, VIC 3010, Australia}

\author{A. Stacey}
\affiliation{Centre for Quantum Computation and Communication Technology, School of Physics, University of Melbourne, Parkville, VIC 3010, Australia}	
\affiliation{School of Science, RMIT University, Melbourne, VIC 3001, Australia}

\author{B. C. Johnson}
\affiliation{School of Physics, University of Melbourne, Parkville, VIC 3010, Australia}
\affiliation{Centre for Quantum Computation and Communication Technology, School of Physics, University of Melbourne, Parkville, VIC 3010, Australia}

\author{D. A. Broadway}
\affiliation{School of Physics, University of Melbourne, Parkville, VIC 3010, Australia}
\affiliation{Centre for Quantum Computation and Communication Technology, School of Physics, University of Melbourne, Parkville, VIC 3010, Australia}

\author{T. Teraji}
\affiliation{National Institute for Materials Science, Tsukuba, Ibaraki 305-0044, Japan}

\author{D. A. Simpson}
\affiliation{School of Physics, University of Melbourne, Parkville, VIC 3010, Australia}

\author{J.-P. Tetienne}
\email{jtetienne@unimelb.edu.au}
\affiliation{School of Physics, University of Melbourne, Parkville, VIC 3010, Australia}
\affiliation{Centre for Quantum Computation and Communication Technology, School of Physics, University of Melbourne, Parkville, VIC 3010, Australia}	

\author{L. C. L. Hollenberg}
\email{lloydch@unimelb.edu.au}
\affiliation{School of Physics, University of Melbourne, Parkville, VIC 3010, Australia}
\affiliation{Centre for Quantum Computation and Communication Technology, School of Physics, University of Melbourne, Parkville, VIC 3010, Australia}

\begin{abstract}

Thin layers of near-surface nitrogen-vacancy (NV) defects in diamond substrates are the workhorse of NV-based widefield magnetic microscopy, which has applications in physics, geology and biology. Several methods exist to create such NV layers, which generally involve incorporating nitrogen atoms (N) and vacancies (V) into the diamond through growth and/or irradiation. While there have been detailed studies of individual methods, a direct side-by-side experimental comparison of the resulting magnetic sensitivities is still missing. Here we characterise, at room and cryogenic temperatures, $\approx100$ nm thick NV layers fabricated via three different methods: 1) low-energy carbon irradiation of N-rich high-pressure high-temperature (HPHT) diamond, 2) carbon irradiation of $\delta$-doped chemical vapour deposition (CVD) diamond, 3) low-energy N$^+$ or CN$^-$ implantation into N-free CVD diamond. Despite significant variability within each method, we find that the best HPHT samples yield similar magnetic sensitivities (within a factor 2 on average) to our $\delta$-doped samples, of $<2$~$\mu$T Hz$^{-1/2}$ for DC magnetic fields and $<100$~nT Hz$^{-1/2}$ for AC fields (for a $400$~nm~$\times~400$~nm pixel), while the N$^+$ and CN$^-$ implanted samples exhibit an inferior sensitivity by a factor 2-5, at both room and low temperature. We also examine the crystal lattice strain caused by the respective methods and discuss the implications this has for widefield NV imaging. The pros and cons of each method, and potential future improvements, are discussed. This study highlights that low-energy irradiation of HPHT diamond, despite its relative simplicity and low cost, is a competitive method to create thin NV layers for widefield magnetic imaging.      

\end{abstract}

\maketitle 

\section{Introduction}

Widefield imaging based on ensembles of nitrogen-vacancy (NV) defects in diamond~\cite{Doherty2013,Rondin2014,Steinert2010,Pham2011,Chipaux2015,Simpson2016,Glenn2015}, also known as quantum diamond microscopy, is a promising tool with applications in condensed matter physics~\cite{Tetienne2017,Casola2018,Broadway2018c}, geology~\cite{Fu2014,Glenn2017} and biology~\cite{LeSage2013,Glenn2015}. Common to all these applications is the need for a dense layer of NV defects near the surface of a single crystal diamond substrate. Depending on the exact application, the optimal thickness $t$ of this layer can vary from $t\lesssim10$~nm to detect mesoscopic fluctuating signals e.g. from paramagnetic ions in solution~\cite{Steinert2013,DeVience2015,Simpson2017}, to several $\mu$m for imaging of geological samples over mm-scale fields of view~\cite{Glenn2017}. 

For applications related to magnetic imaging, it is desirable to achieve diffraction limited resolution ($\Delta x\approx300$~nm) and so the layer thickness should satisfy $t <\Delta x$. This regime is also convenient because it allows the NV layer to be treated as a 2D layer when reconstructing the current density or magnetisation of the sample~\cite{Tetienne2017,Broadway2020,Ku2019}, and it preserves the ability to image the sample via NV photoluminescence (PL) quenching or laser interference effects~\cite{Tisler2013,Tetienne2019,Lillie2019}. A thickness of order $t \approx 100$~nm is found to be optimal as it allows one to maximise the number of NVs without impacting the spatial resolution, while also permitting the addition of a spacing layer (e.g. $t_{\rm sp}=100$~nm of Al$_2$O$_3$~\cite{Broadway2020}) grown on the diamond surface, by satisfying $t+t_{\rm sp}<\Delta x$. The use of such a spacing layer is required to mitigate proximity-induced artefacts arising from strongly magnetic samples \cite{Tetienne2018b}.

Several approaches exist to create such thin near-surface NV layers \cite{Smith2019}. To localise NV centres near the surface one must restrict the creation of vacancies or incorporation of nitrogen (but not necessarily both) to the near-surface region. The N impurities can be introduced during the diamond synthesis, either via the high-pressure high-temperature (HPHT) method or via chemical vapour deposition (CVD) \cite{Achard2020}, or via ion implantation of an N-containing species~\cite{Pezzagna2011a}. The vacancies are typically introduced by irradiation with electrons, protons or ions~\cite{Acosta2009}. The diamond is then annealed at $800-1200^\circ$C in a vacuum environment to allow the vacancies to diffuse and form NV centres~\cite{Deak2014}. The sensitivity of the resulting NV layer depends not only on the number of NVs but also on their ``quality'', in particular their charge state (only NV$^-$ is useful for quantum sensing applications, whereas the co-presence of NV$^0$ produces background PL reducing the sensitivity) and spin properties (coherence times $T_2^*$ and $T_2$, optical contrast)~\cite{Rondin2014,Tetienne2018a,Barry2019}. These factors depend on the presence of other defects in the lattice (non converted N impurities or other growth- or irradiation-induced defects) and can vary significantly between different NV creation methods, motivating a comparative study of the resulting magnetic sensitivities. 

Assuming shot noise limited measurements, the sensitivity to DC magnetic fields with the pulsed optically detected magnetic resonance (ODMR) protocol is approximately given by \cite{Dreau2011,Rondin2014,Barry2019}
\begin{eqnarray} \label{Eq:eta_dc}
\eta_{\rm dc}\equiv\frac{1}{\gamma_e {\cal C}T_2^*\sqrt{\alpha {\cal R}}}
\end{eqnarray}
where $\gamma_e\approx2.8$~MHz/G is the electron gyromagnetic ratio, ${\cal C}\ll 1$ is the spin contrast, $T_2^*$ is the spin dephasing time, ${\cal R}$ is the photon count rate under continuous wave (CW) excitation, and $\alpha=\frac{t_R}{t_R+t_D+T_2^*}$ is the readout duty cycle where $t_R$ and $t_D$ are the readout time and dead time per cycle, respectively. As ${\cal R}$ will scale with the NV density, itself usually proportional to the density of other unwanted paramagetic defects such as the substitutional nitrogen defect, it competes with the ensemble's $T_2^*$ and it is not a priori clear which attribute to prioritise. In addition, the sensitivity for given parameters depends on the experimental conditions through $t_R$ and $t_D$. For widefield imaging, the low laser intensities used require a relatively long $t_R$ ($\approx$~10~$\mu$s), meaning $\alpha\approx 1$ and Equation \ref{Eq:eta_dc} retains a strong $T_2^*$ dependence. This contrasts with the confocal microscopy case, where often $\alpha \approx t_R/T_2^*$. The sensitivity to AC or fluctuating magnetic fields, $\eta_{\rm ac}$, has a similar expression where $T_2^*$ is replaced by $T_2$~\cite{Taylor2008,Rondin2014,Barry2019}, which similarly scales inversely to $\cal R$.

In this work, we focus on three commonly employed methods to create thin NV layers:
\begin{itemize}
\item Method~1: low-energy irradiation of N-rich HPHT diamond;
\item Method~2: Irradiation of $\delta$-doped CVD diamond;
\item Method~3: low-energy implantation of N-containing species into N-free CVD diamond.
\end{itemize}
Method~1 aims to create vacancies confined to a distance $\approx t$ from the surface of an HPHT diamond that contains a high density of N throughout the bulk, typically of the order of $[\text{N}]\approx100$~ppm. This method has been previously demonstrated using $\sim30$~keV helium ions (He$^+$), producing NV layers with $t\sim100-200$~nm~\cite{Huang2013,McCloskey2014,Fescenko2019}. Here we employ C$^-$ ions, which is a readily available species in low energy negative ion sources. One advantage over He$^+$ is the absence of He-related defects in the final NV layer.  Method~2 replaces the HPHT substrate with a CVD diamond in which nitrogen was selectively incorporated during the growth so as to form a $\delta$-doped layer of desired N density and thickness, on an otherwise N-free substrate~\cite{Achard2020,Ohno2012}. The vacancies can then be created by low-energy ion implantation (here we also use C$^-$)~\cite{Ohno2014,FavarodeOliveira2016,Kleinsasser2016}, although higher energy particles could also be used since the vacancies do not have to be confined in depth, for instance NVs have been formed using 200 keV electrons from a transmission electron microscope~\cite{Kim2012,McLellan2016}. Finally, Method 3 employs low-energy implantation of an N-containing species (here we consider N$^+$ and CN$^-$) to simultaneously incorporate both N and V at the desired depth into an otherwise N-free diamond~\cite{Pezzagna2010,Toyli2010,Spinicelli2011}. Despite being a very popular method for producing NV sensing layers, this approach creates unnecessary damage to the diamond because of the large number of vacancies created for each N~\cite{Antonov2014,DeOliveira2017}, which is especially problematic for dense layers of near-surface NVs~\cite{Tetienne2018a}. In contrast, in Methods 1 and 2 the irradiation fluence can be adjusted to reach a ratio of created vacancies to the number of N impurities present in the diamond optimal for maximising NV yield, making them better candidates to reach the regime in which $T_2^*$ is limited only by the residual N impurities~\cite{Barry2019,Bauch2019}. In practice, this condition is likely to be given by a combination of the resultant Fermi level position as well as the nitrogen and vacancy concentrations, and thus optimising implant parameters is expected to require a level of fine tuning beyond the scope of this work. For simplicity, we focus on the regime where the number of vacancies created by the implantation process is commensurate with the expected nitrogen concentration. 

While these techniques have individually been the subject of previous works~\cite{Huang2013,McCloskey2014,Fescenko2019,Ohno2012,Ohno2014,FavarodeOliveira2016,Kleinsasser2016,Kim2012,McLellan2016,Pezzagna2010,Toyli2010,Spinicelli2011,Antonov2014,DeOliveira2017}, the large variability in the experimental methods used to assess them make a reliable comparison difficult. In particular, the quantities ${\cal C}$, $\alpha$ and ${\cal R}$ in Eq.~(\ref{Eq:eta_dc}) all depend on the specifics of the optical setup used to measure them, and can vary drastically between confocal and widefield measurements. In widefield magnetic imaging, the laser intensities are typically much smaller than in confocal measurements, and so the per-NV sensitivity may appear smaller, which is a price to pay to access a large field of view. Thus, the purpose of the present work is to provide a side-by-side comparison of the sensitivities ($\eta_{\rm dc}$ and $\eta_{\rm ac}$) obtained with the three NV creation methods outlined above, while keeping as many experimental parameters fixed as possible. We stress that, given the large number of experimental variables for each method, we do not seek to optimise them. Instead, our aim is to generate a baseline of the performances that can be achieved with the different methods without advanced optimisation, so as to help users to best choose the method that suits their capabilities and application, and to guide future efforts in optimising a particular method.

The paper is organised as follows. In Sec.~\ref{sec:methods}, we describe the experimental procedures to form the NV layers studied (\ref{sec:samples}) and to characterise them (\ref{sec:meas}). The results of these measurements are presented in Sec.~\ref{sec:results}, followed by a discussion (\ref{sec:discussion}) and conclusion (\ref{sec:conclusion}).

\section{Methods} \label{sec:methods}

\subsection{Samples} \label{sec:samples}

%Refer to Table~\ref{Table:samples} and detail the full methods. Discuss polished vs overgrown which was found to make no difference for dense ensembles~\cite{Tetienne2018a}.
\begin{figure*}
\includegraphics[width=0.7\textwidth]{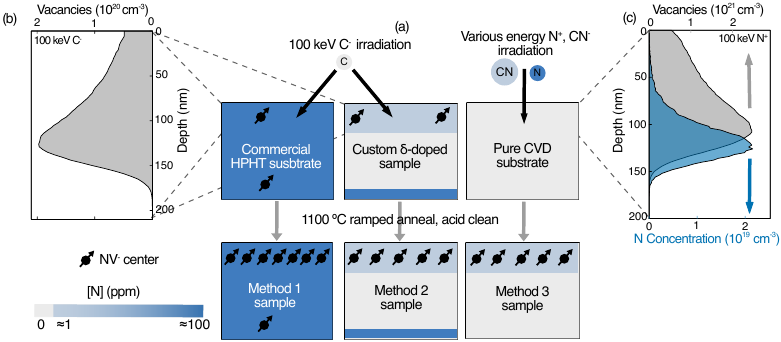}
\centering
\caption{\textbf{Overview of the NV layer formation methods investigated:} \textbf{a)} Schematic illustrating sample production. From left to right: Method 1, carbon irradiation of high-N, low-NV HPHT diamond to create a localised NV-rich layer; carbon irradiation of $\delta$-doped CVD layer to improve NV conversion; implantation of a pure CVD substrate with a chosen nitrogen-containing species to incorporate nitrogen into the crystal at the cost of extra damage by vacancy over-creation. The colour scale used and relative NV densities are schematic only. \textbf{b)} and \textbf{c)} are SRIM simulations for the C$^-$ and N$^+$ implantation cases respectively, showing depth distributions of vacancies produced (both) and implanted ions (Method 3 only) produced by representative implantation procedures. }
\label{fig:schematic}
\end{figure*}
The sample creation methods used for this study are detailed in Table \ref{Table:samples} and depicted schematically in Figure \ref{fig:schematic}(a). Method 1 samples (referred to as HPHT samples hereafter) are as-received commercially available (Delaware Diamond Knives) HPHT diamonds polished to a surface roughness $<$5~nm Ra with specified nitrogen concentration of $<$~200~ppm. Despite the high initial nitrogen concentration, these samples showed limited NV fluorescence prior to implantation indicating poor N to NV conversion. C$^-$ irradiation with doses between $10^{11}$ and $10^{13}$~cm$^{-3}$, chosen to produce a number of vacancies comparable to the expected nitrogen concentration, resulted in significant PL enhancement (by several orders of magnitude). Samples were held on the implanter stage at a 7$\degree$ tilt angle and 10$\degree$ rotation to minimise ion channelling, and the chamber pressure during implantation was $\leq 10^{-6}$~mbar.  The implant energy of 100~keV was chosen to confine vacancies produced to a $\sim$100~nm layer as shown by Figure \ref{fig:schematic}(b), which is a simulation produced using the Stopping and Range of Ions in Matter (SRIM) software package.

Samples fabricated by Method 2 (hereafter referred to as $\delta$-doped samples) were created via CVD growth over a commercial type Ib HPHT substrate (Element Six). An oxygen etch was performed to remove polishing damage and maximise subsequent overgrowth quality \cite{Tallaire2004}. An electronic grade, N-free layer of a thickness order 10~$\mu$m was then grown to buffer the grown NVs from defects in the substrate and to reduce the PL contribution from the substrate. The $\delta$-doped layer was then grown on top during a short growth designed to create a film between 100 and 200~nm thick and with a level of nitrogen incorporated set by the proportion of nitrogen present in the plasma during growth (varied between samples). The $\delta$-doped layer was grown using a low microwave power of 750~W and under a low pressure of 25~Torr to promote slow, high quality crystal growth \cite{Ohno2012}. The proportion of methane (natural isotopic proportions) in the gas mixture was $\approx$~1\%. The CVD reactor used to grow the NV layers is multi-purpose and hence not specifically optimised for low density incorporation of growth defects, so the grown samples are not expected to represent benchmark quality. Instead, these samples are intended to represent ``standard" and relatively easy to produce $\delta$-doped diamonds. Prior to implantation the doped layer exhibited some fluorescence but it is expected that less than 0.5\% of incorporated nitrogen forms NV centres in the as-grown diamond \cite{Edmonds2012}. This PL was increased by over an order of magnitude on average by 100~keV, $10^{11}$-$10^{12}$~cm$^{-2}$ carbon irradiation, indicating an improved NV yield. The implant parameters were chosen to approximately match the vacancy production to the expected N density and depth profile of the doped layer. 

Method 3 samples were made from electronic-grade single-crystal CVD diamonds (purchased from Delaware Diamond Knives) and then overgrown with 2~$\mu$m of  $^{12}$C-enriched (99.95\%) electronic-grade CVD diamond \cite{Teraji2015}, although this step is not normally required for this method. The substrates were then implanted using $^{15}$N$^+$ (30-100~keV, 4$\times 10^{13}$-1$\times 10^{14}$~cm$^{-2}$) or $^{12}$C$^{14}$N$^{-}$ (54~keV, 2$\times 10^{12}$~cm$^{-2}$) ions (hereafter referred to as N$^+$ and CN$^-$ implants), which were chosen as representative nitrogen implant species as they are commonly available at most low energy ion implantation facilities. Figure \ref{fig:schematic}(c) shows this method produces an overlapping region of implanted ions and vacancies but without the ability to tune densities (here the peak vacancy density is two orders of magnitude larger than the peak N density according to SRIM). 

%Diamonds were implanted with various ions at different energies and doses, summarised in Table \ref{Table:samples} (full plots to be shown in appendix/SI). Samples were held on the implanter stage at a 7$\degree$ tilt angle and 10$\degree$ rotation to minimise ion channelling. The chamber pressure during implantation was $\leq 10^6$~mbar. Figure \ref{fig:schematic} includes depth profiles of both implant ions and vacancies created for the range of energies that form this study, simulated using the Stopping and Range of Ions in Matter (SRIM) software package. The energies used were chosen so as to create near-surface NV layers that satisfied the conditions of allowing diffraction-limited imaging and being able to be modelled as a 2D layer for field reconstruction.  
 
Following implantation, all samples were annealed in a vacuum of $\sim 10^{-5}$~Torr using a ramp sequence culminating at 1100 $\degree$C to form NV centres and repair part of the irradiation-induced damage \cite{Tetienne2018a}.
%6~h at 400~$\degree$C, 2~h ramp to 800~$\degree$C, 6~h at 800~$\degree$C, 2~h ramp to 1100~$\degree$C, 2~h at 1100~$\degree$C, 2~h ramp to room temperature. 
The samples were then cleaned for 15 minutes in a boiling mixture of sulphuric and nitric acid prior to measurement. This is expected to standardise the surface chemistry across all samples, avoiding the impact of the surface on our comparative measurements. Additional sample production details are given in Appendix \ref{sampledetails}.

\begin{table*}[hbt!]
\begin{tabular}{|c|c|c|c|c|c|c|c|c|}
\hline
Sample & Substrate type & Surface & Implanted & Energy & Dose & Projected & Estimated NV density & Estimated NV yield \\
name &  & finish & species & (keV) & (ions/cm$^2$) & range (nm) & (NV/$\mu$m$^2$) & (\%) \\ \hline
HPHT-1 & HPHT & Polished & $^{12}$C$^-$ & 100 & $5\times10^{11}$ & 142 &   1.6$\times10^5$ & 0.90 \\
HPHT-2 & HPHT & Polished & $^{12}$C$^-$ & 100 & $1\times10^{12}$ & 142 &   1.5$\times10^5$ & 0.85 \\
HPHT-3 & HPHT & Polished & $^{12}$C$^-$ & 100 & $1\times10^{13}$ & 142 &   4.6$\times10^5$ & 2.60 \\
HPHT-4 & HPHT & Polished & $^{12}$C$^-$ & 100 & $1\times10^{13}$ & 142 &   1.3$\times10^5$ & 0.73 \\
HPHT-5 & HPHT & Polished & $^{12}$C$^-$ & 100 & $1\times10^{13}$ & 142 &   4.7$\times10^5$ &  2.66\\
HPHT-6 & HPHT & Polished & $^{12}$C$^-$ & 100 & $2\times10^{11}$ & 142 &   4.5$\times10^5$ &  2.54\\
HPHT-7 & HPHT & Polished & $^{12}$C$^-$ & 100 & $1\times10^{13}$ & 142 &   4.2$\times10^5$ &  2.37\\
$\delta$-1 & CVD ($\delta$-doped) & As grown & $^{12}$C$^-$ & 100 & $1\times10^{12}$ & 142 &  2.6$\times10^4$ &  \\
$\delta$-2 & CVD ($\delta$-doped) & As grown & $^{12}$C$^-$ & 100 & $1\times10^{12}$ & 142 &   4.8$\times10^4$ &  \\
$\delta$-3 & CVD ($\delta$-doped) & As grown & $^{12}$C$^-$ & 100 & $1\times10^{11}$ & 142 &   1.5$\times10^3$ &  \\
$\delta$-4 & CVD ($\delta$-doped) & As grown & $^{12}$C$^-$ & 100 & $1\times10^{11}$ & 142 &   2.6$\times10^3$ &  \\
$\delta$-5 & CVD ($\delta$-doped) & As grown & $^{12}$C$^-$ & 100 & $1\times10^{12}$ & 142 &   2.9$\times10^4$ &  \\
N-1 & CVD & As grown & $^{15}$N$^+$ & 30 & $4\times10^{13}$ & 42 &   1.7$\times10^4$ & 4.12 \\
N-2 & CVD  & As grown & $^{15}$N$^+$ & 30 & $4\times10^{13}$ & 42 &   2.6$\times10^4$ & 6.43 \\
N-3 & CVD & As grown & $^{15}$N$^+$ & 100 & $4 \times10^{13}$ & 122 &   2.6$\times10^4$ & 4.54 \\
N-4 & CVD  & As grown & $^{15}$N$^+$ & 100 & $5\times10^{13}$ & 122 &   2.9$\times10^4$ & 5.81 \\
N-5 & CVD  & As grown & $^{15}$N$^+$ & 100 & $1\times10^{14}$ & 122 &   2.0$\times10^4$ & 2.00 \\
CN-1 & CVD & As grown & $^{12}$C$^{14}$N$^-$ & 54 & $2\times10^{13}$ & 42 &   9.1$\times10^3$ & 4.55 \\
CN-2 & CVD  & As grown & $^{12}$C$^{14}$N$^-$ & 54 & $2\times10^{13}$ & 42 &   7.7$\times10^3$ & 3.87 \\
CN-3 & CVD  & As grown & $^{12}$C$^{14}$N$^-$ & 54 & $2\times10^{13}$ & 42 &   7.2$\times10^3$ & 3.61 \\ 
\hline
\end{tabular}
\caption{\textbf{List of diamond samples used in this study.} Samples differ in (from left to right in the table) their diamond growth method, whether overgrowth took place after substrate polishing, and the implant species, energy, and dose chosen. NV density estimated by comparing fluorescence to that given by a single NV on a confocal microscope. Yields quoted are calculated by comparing the inferred NV content with the known N fluence for the N implants and taking an estimated N concentration of 100~ppm for HPHT samples. No precise measure of doped nitrogen content is available for $\delta$-doped layers.}
\label{Table:samples}
\end{table*}

\subsection{Measurements} \label{sec:meas}

The sensitivity measurements for all samples were carried out on a purpose-built widefield microscope described in detail elsewhere \cite{Tetienne2018a, Broadway2018c}. Microwave delivery is facilitated by a gold resonator deposited on a glass cover slip, onto which diamond chips are mounted NV face-down. A permanent magnet aligned with one family of NV axes provides the bias field required to Zeeman split the NV resonances. NV centres are excited by a 532~nm laser focussed to the back aperture of an oil-immersion microscope objective (Nikon 40x NA = 1.3), where the laser power was measured to be 220~mW. The resulting NV fluorescence was filtered (660-735 nm) and imaged on a scientific complementary metal oxide semiconductor (sCMOS) camera for processing. The measurements were taken over a 50 $\mu$m $\times$ 50 $\mu$m field of view as this was the region over which the laser intensity was roughly uniform ($\approx$~3~kW/cm$^2$). This region, typical for widefield imaging experiments, contains $10^6$-$10^7$ NVs (depending on the sample) which we average over in our spin measurements. A readout pulse time of $t_R = 5$~$\mu$s, chosen to reach a compromise between ensemble initialisation and readout contrast, was held constant for all measurements. A dead time of $t_D = 1.5$~$\mu$s was introduced following each laser pulse to allow relaxation from the metastable singlet state to the ground state. All room temperature measurements, with the exception of $T_2$, were conducted under a weak bias field of 60~G. $T_2$ measurements were undertaken at a higher field of 475~G to ensure that, for the samples containing a natural abundance of $^{13}$C, the resulting revivals were closely spaced enough to allow accurate extraction of the decay envelope. The microwave power was held to give a constant $\pi$~time of 40~ns for all spin measurements. To remove common mode laser noise (where applicable), pulse sequences were normalised using a reference measurement taken with respect to the $\ket{-1}$ basis rather than the usual $\ket{0}$ by using a $\pi$ pulse prior to readout. 

To characterise the samples, we measure the quantities present in Equation \ref{Eq:eta_dc}. The photon count rate $\cal R$ was averaged over the field of view under CW laser excitation over a consistent camera exposure time of 30~ms, and then normalised in time and area. Estimated NV densities listed in Table \ref{Table:samples} were inferred from $\cal R$ using a scaling factor obtained by comparing the fluorescence collected on a confocal microscope for selected samples to that from a reference single NV under identical conditions.  The spin contrast is obtained by coherently driving the NV ensemble, with ${\cal C}$ taken as the peak-to-peak amplitude of the Rabi oscillations when fit to a damped sinusoid. $T_2^*$ is measured using the Ramsey pulse sequence and the $T_2$ is measured first with a Hahn echo sequence and then with Carr-Purcell-Meiboom-Gill (CPMG) sequences with variable numbers of $\pi$ pulses. We also measure the spin relaxation time $T_1$ which represents an upper limit to ensemble coherence times and, due to its sensitivity to resonant magnetic fluctuations, forms the basis of many sensing experiments \cite{Simpson2016, McCoey2020}. It is measured simply as the decay of the $\ket{0}$ state under dark evolution. Further details on the measurements and data analysis are given in Appendix \ref{app: measurements} and full results for each sample are given in Table \ref{Table:results} (Appendix \ref{app:results}).

These measurements were repeated at lower temperatures (down to 4~K) for representative samples on a similar widefield NV microscope within a close-cycle cryostat \cite{Lillie2020}. To minimise sample heating, a lower laser power was used for these measurements, which required a longer readout time of $t_R=10$~$\mu$s to be used to compensate for the reduced rate of ensemble initialisation.

\section{Results} \label{sec:results}

\subsection{Basic spin properties}
\label{sec: results1}

\begin{figure}
\centering
	\includegraphics[width=0.45\textwidth]{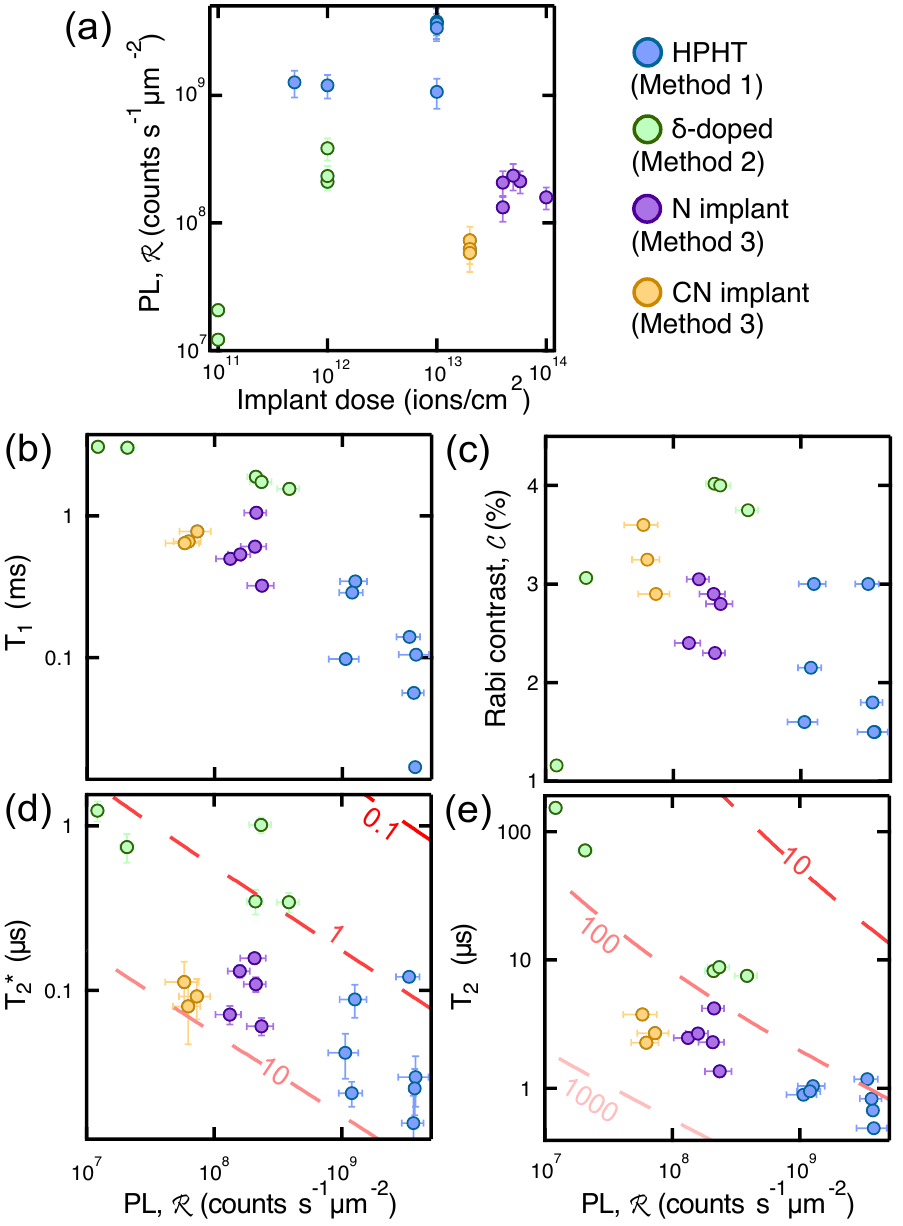}
	\caption{\textbf{Sample characterisation at room temperature:} Measured variation of \textbf{a)} $\cal R$ with implant dose; and \textbf{b)} $T_1$, \textbf{c)} \ Rabi contrast, \textbf{d)} free induction decay time $T_2^*$, and \textbf{e)} Hahn echo $T_2$ with $\cal R$. All measurements made at room temperature and at 60~G except for the Hahn echo, which was conducted at 475~G to reduce the impact of $^{13}$C revivals on the fits. Red lines in \textbf{d)} and \textbf{e)} denote lines of constant per 400~nm~$\times$~400~nm DC and AC sensitivity (in $\mu$T~Hz$^{-1/2}$ and nT~Hz$^{-1/2}$ respectively) respectively at a constant Rabi contrast of 3\% to guide the eye. The error bars indicate one standard deviation for $\cal R$ and the standard fitting errors for the spin properties. Where not visible, error bars are smaller than the marker size.}  
	\label{Fig: coherence}
\end{figure}

To benchmark sample quality we first consider the basic NV spin properties at room temperature, summarising the results in Figure \ref{Fig: coherence}. Focussing solely on PL for now, we plot the photon count rate $\cal R$ against the implantation doses used for all the samples studied in Figure \ref{Fig: coherence}(a), with the different methods separated by colour. The HPHT samples are the brightest, which is to be expected given their high initial N concentration ($\approx$100~ppm). Method 3 samples exhibit fluorescence within an order of magnitude of each other, mostly in line with the spread in implantation doses as reflected by the fairly constant conversion ratios listed in Table \ref{Table:samples}. These samples are less fluorescent than the HPHT samples and brightest $\delta$-doped samples despite higher implant doses because the NV production is limited by the nitrogen incorporated during the implant process rather than the vacancy production. $\delta$-doped samples feature the widest spread in brightness due to nitrogen concentration in the CVD gas mixture being varied over several orders of magnitude, although we do not expect a direct conversion from nitrogen concentration in the reactor gas mixture to that incorporated into the crystal. Indeed, the similar PL of the three brightest $\delta$-doped samples suggests we may be close to saturating the nitrogen concentration under our growth conditions. 

In Figure \ref{Fig: coherence}(b)-(e) the results of the spin measurements described above are plotted against the $\cal R$ as this is the parameter from Equation \ref{Eq:eta_dc} expected to vary most straightforwardly between samples, being proportional to NV density. Figure \ref{Fig: coherence}(b) plots $T_1$ against $\cal R$ and we see that the $T_1$ time varies significantly between the different methods and is inversely correlated to PL. $T_1$ ranges from $\sim$3~ms in the lowest density $\delta$-doped samples, which we assume to be a nearly phonon-limited value, to $\lesssim$100~$\mu$s for the highest PL HPHT samples. We attribute this dramatic reduction in $T_1$ proportional to the increase in nitrogen defects both to the general increase in magnetic noise and to electron tunnelling to nearby defects that sees the NV$^-$ ionise to NV$^0$ during dark evolution~\cite{Bluvstein2019,Manson2018}.

Figure \ref{Fig: coherence}(c) shows that the Rabi contrast $\cal C$ also appears to scale inversely with PL, although the variation is milder with most samples lying in the range of 2-4\%. As in the $T_1$ case, the availability of extra tunnelling sites in the higher density samples is likely responsible for the reduced measurement contrast. The lowest density $\delta$-doped sample is an exception to this trend, likely because this sample contains too low a nitrogen density to sustain a high NV$^-$ yield in the presence of signifcant surface-induced band bending \cite{Stacey2019}. 

The inverse correlation between free induction decay time $T_2^*$ and $\cal R$ (Figure \ref{Fig: coherence}(d)) is expected as the increased defect density associated with brighter samples gives a higher level of magnetic noise, which reduces $T_2^*$. To see how this impacts measurement sensitivity, we plot constant sensitivity (calculated for a 400~nm~$\times$~400~nm ``pixel", representing approximately diffraction-limited spatial resolution) curves (red dashed lines) assuming 3\% contrast in Equation \ref{Eq:eta_dc}. The data points roughly follow these curves, indicating that $T_2^*$ and $\cal R$ offset one another to a large degree, consistent with previous studies \cite{Tetienne2018a}. However, there is a large spread in this data and additional trends within and between the methods are worth noting. The largest variation in NV quality relative to PL (vertical spread in figures) is present in the HPHT samples (with $T_2^*$ ranging from 15 to 120~ns). This variation is due to this method's reliance on a purchased substrate not produced to specification, as well as the variation in defect incorporation in different HPHT growth sectors. The best HPHT samples, however, appear to outperform the nitrogen implanted samples, as do all of the $\delta$-doped samples. This confirms that the additional damage associated with nitrogen implantation produces a source of dephasing that is significant alongside the contribution from the nitrogen spin bath. This is further evidenced by the samples implanted with the larger CN$^-$ ion exhibiting comparable coherence times to the N$^+$ implants, but with much lower PL. In other words, the extra damage caused per ion (a 54~keV CN$^-$ implant produces $65$\% and $25$\% more vacancies than 30~keV and 100~keV N$^+$ implants respectively) amounts to a similar total paramagnetic defect density for a lower nitrogen concentration.

Figure \ref{Fig: coherence}(e) plots Hahn echo $T_2$ against $\cal R$, again with constant sensitivity curves to guide the eye. The trends are similar to the $T_2^*$ case, but with a reduction in spread due to the success of the Hahn echo sequence at removing quasi static magnetic noise contributions. 

\begin{figure}
\includegraphics[width=0.45\textwidth]{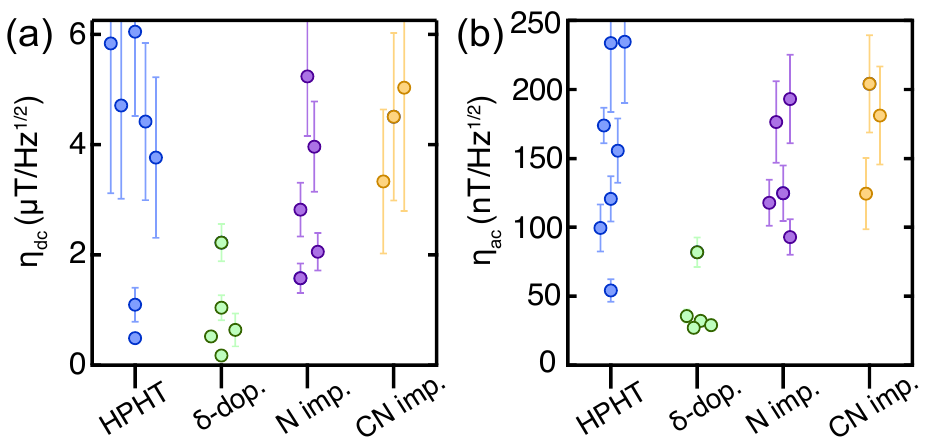}
\centering
\caption{\textbf{Sensitivity vs NV creation method:} Values of \textbf{a)} DC and \textbf{b)} AC magnetic sensitivities from Equation \ref{Eq:eta_dc} (using $T_2^*$ and $T_2$ respectively) for a 400~nm $\times$ 400~nm pixel. Error bars obtained using standard propagation using measurement errors detailed in Figure \ref{Fig: coherence}. Sensitivities obtained on a setup with $t_R=5$~$\mu$s, $t_D=1.5$~$\mu$s, and a laser intensity of 3~kW/cm$^2$.}
\label{fig: sensitivity}
\end{figure}

To better analyse the trends in sensitivity versus fabrication method, we plot the calculated DC and AC sensitivities (using actual Rabi contrast values for each sample) in Figure \ref{fig: sensitivity}. We note that variance in NV layer thickness of up to a factor of 2 translates to a scaling in sensitivity of up to $\sqrt{2}$.  It is also worth noting that the laser power used is far below optical saturation and so sensitivity could be improved by a further 10 times or more for any given sample by focussing the laser and increasing the counts received per unit area by a factor of 100 (at the cost of a decreased field of view). Nevertheless, the trends qualitatively noted earlier are confirmed: the $\delta$-doped samples emerge as the highest quality, with the mean magnetic sensitivity for this method being around a factor of 3 better than that of the others. The HPHT series exhibits the largest amount of spread (almost 100~\% of the mean), but the best HPHT samples provide comparable sensitivity to the $\delta$-doped samples.
This spread in sensitivity is accompanied by variability in the amount of fluorescence a given implantation dose promotes for HPHT samples, which is indicative of the variability in nitrogen concentration in the purchased crystal. This lack of foreknowledge can be compounded by a failure to correctly tune the vacancy production of the chosen implantation procedure, and hence a failure to optimise the process. Over-implanting will produce unnecessary damage and under-implanting will result in a suboptimal NV:N conversion ratio, and in both cases this will be reflected by a decrease in sensitivity. This suggests that a way to improve the sensitivity of these samples could be to determine the N density prior to implantation and adjust implant parameters accordingly. The HPHT samples already achieving the best sensitivities are likely those currently best optimised, and so it is reasonable to expect refinement of the process to result in this level of sensitivity being achieved (or improved upon) by a majority of HPHT samples in future. There is also potential to improve the $\delta$-doped samples' sensitivity in this way as the N content is also initially unknown in this case, but this improvement does not exist for the implantation of an N-free substrate as there is no way to alter the ratio of vacancy production to N implanted at a given energy. 

These results indicate that the simple and relatively inexpensive Method 1 can offer competitive magnetic sensitivity in widefield NV imaging, with large potential for improvement following further optimisation. 

\subsection{Extended dynamical decoupling}
\label{sec: cpmg}

In sensing AC fields, it is common to use more advanced dynamical decoupling techniques than the Hahn echo pulse sequence to decouple from the spin environment more efficiently, access higher frequencies, and improve spectral selectivity \cite{Naydenov2011,DeLange2011}. To benchmark our samples for use in this regime, we test their performance under CPMG sequences of increasing length, characterised by the number of $\pi$ pulses $N$.
 
\begin{figure}
\includegraphics[width=0.45\textwidth]{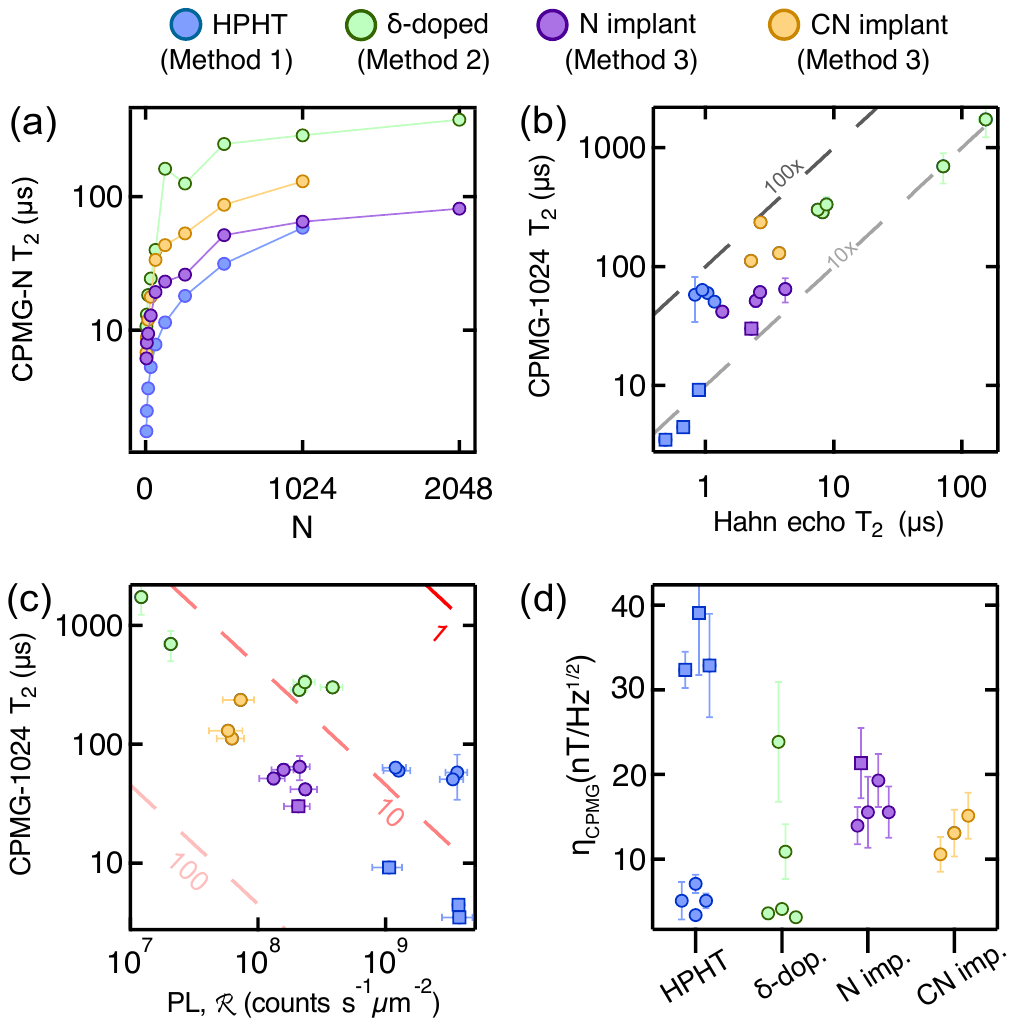}
\centering
\caption{\textbf{Extended decoupling characterisation:} \textbf{a)} $T_2$ scaling of selected representative samples of each sample type as a function of the number of pulses $N$ in the CPMG-$N$ sequence. Data displayed collected from samples $\delta$-1 (green), CN-3 (yellow), N-3 (purple), and HPHT-6 (blue). \textbf{b)} Extension of $T_2$ from $N$=1 (Hahn echo) to $N$=1024, with grey lines showing 10-fold and 100-fold improvement.  \textbf{c)} Scaling of CPMG-1024 $T_2$ with $\cal R$, with iso-sensitivity curves in nTHz$^{-1/2}$. \textbf{d)} AC sensitivities calculated using Equation \ref{Eq:eta_dc} (using the CPMG-1024 $T_2$), grouped by method. Data points marked by squares in panels b)-d) belong to samples that could not be extended beyond $N\approx 256$ and see only modest improvement over Hahn echo values.}
\label{fig: cpmg}
\end{figure}

Figure \ref{fig: cpmg}(a) plots the CPMG $T_2$ as a function of $N$ for representative samples from each method. A clear extension of $T_2$ is evident in all cases, with saturation above $N\approx1024$. Fitting the $T_2$ extension to a power law, $T_2(N) = T_2(1)N^s$, gives values of $s$ between 0.4 and 0.8 across all samples, with the majority of samples exhibiting values below the theoretical $s=2/3$ for a dilute electron spin bath \cite{DeLange2010}. In most cases, the maximum attainable $T_2$ fell short of the $T_2 = T_1/2$ limit previously observed for bulk ensembles \cite{Bar-Gill2013}. Previous work on shallow ($\leq 20$~nm) single NVs saw values of $s<2/3$ and saturated $T_2<T_1/2$ \cite{Romach2015} and as our layers exist in an intermediate depth regime between this and the bulk case, our results are broadly consistent with these studies. No trends specific to the methods of layer creation were obvious so we leave analysis of decoupling scaling of ensembles like these to future work. 

To assess the efficiency of the decoupling across samples, we plot the CPMG-1024 vs Hahn echo $T_2$ for all samples in Figure \ref{fig: cpmg}(b). The enhancement varies between 10 and 100 and is greatest for the CN$^-$ implants and some HPHT samples. The HPHT samples that see the least extension could not be extended to $N=1024$ due to the finite pulse length (see below), while N implants saw less $T_2$ extension than the other methods despite the majority supporting this number of pulses. The CN$^-$ implants undergo greater $T_2$ extension than the N$^+$ implants, indicating that the decoupling is successful in mitigating the effects of the extra vacancy production. 

Comparing CPMG-1024 $T_2$ to $\cal R$ we see (Figure \ref{fig: cpmg}(c)), as before, that the data follows constant sensitivity curves but in two apparently separate bands. The HPHT samples that were successfully decoupled to $N=1024$ display comparable $T_2$ to the N-implants, which are much less fluorescent, and lie along a similar iso-sensitivity line to the bright $\delta$-doped samples. This is confirmed in Figure \ref{fig: cpmg}(d), which shows the best magnetic sensitivities are obtained by $\delta$-doped and HPHT samples. 

The lowest density $\delta$-doped samples fall slightly behind in sensitivity compared to their DC or Hahn echo AC sensitivities despite achieving $T_2$ values close to $T_1$. We can understand this by viewing Equation \ref{Eq:eta_dc} in the long $T_2$ limit ($T_2 \gg t_R=5$~$\mu$s), in which $\alpha$ tends towards $t_R/T_2$ rather than 1. This shows that as $T_2$ is extended the dependence of the equation on $\cal R$ becomes stronger, and this is part of the reason why the HPHT samples excel in this regime. The overall spread in sensitivity within and between methods (ignoring samples that could not be extended to $N=1024$) is again reduced with additional decoupling, with the Method 3 samples, although still the least sensitive on average, only obtaining worse sensitivities by a factor $\sim$2. 

\iffalse
\begin{figure}
\includegraphics{fig5.pdf}
\centering
\caption{\textbf{Example CPMG series} \textbf{a)} Sample HPHT-6 (HPHT) and \textbf{b)} Sample $\delta$-1 ($\delta$-doped) series are typical of the extreme cases (inset numbers denote corresponding $N$ values): HPHT samples and others with low coherence times successfully decouple but pushing beyond $N=1024$ (and often even up to this point) is made difficult because the finite pulse length used can restrict the gains achievable with extended decoupling as the microwave duty cycle of total time $N\times 40$~ns becomes commensurate with $T_{1\rho}$. Where less significant decoherence occurs during the microwave drive, as in \textbf{b)}, decoupling out to larger $N$ values is successful, allowing measurement of values approaching the true, limiting $T_2$ value. Even for these samples, however, only limited improvements are made by extending $N$ past 1024.}
\label{fig: cpmgseries}
\end{figure}
\fi
We note that a key experimental limitation for this section is the microwave infrastructure that imposes a finite pulse duration (here 40~ns for each $\pi$ pulse). While the idealised instantaneous pulse approximation is valid when only a small number of pulses are applied, as $N$ is increased the microwave duty cycle can become significant and the bulk of dephasing can occur during this driving, limiting the effectiveness of decoupling. This is especially evident for the N-dense HPHT samples and is the reason why the majority of samples reach limiting $T_2$ values far shorter than $T_1$. In principle, faster pulses could see additional coherence extension and the AC sensitivities of the worst-performing HPHT samples fall into line with the others, further incentivising the creation of highly dense NV ensembles, but in practice this is a realistic experimental limitation.

Overall, the conclusions in this sensing regime are much the same as in the previous section, apart from the most optimised HPHT samples becoming even more competitive and the impact of the disadvantages inherent to Method 3 appearing to be reduced. 

\subsection{Crystal strain mapping}
\label{section: strain}

Separate from magnetic sensitivity, another important quantity for widefield magnetic imaging is the crystal strain homogeneity in the NV layer. In particular, strain inhomogeneity can easily be conflated with the variation of a target field (in DC magnetometry), or constitute variable detuning from resonant driving across the field of view (AC measurements). Additional experiments can be made to normalise out these effects, though this is often impractical as it requires addressing multiple transition frequencies \cite{Broadway2019}. This could make an experiment less time efficient or be technologically impossible for high fields, hence it is advantageous to have as little strain variation as possible. 

Our samples vary in their surface finish (polished vs overgrown) and their growth method (CVD vs HPHT), both factors which will contribute to near-surface strain. The samples also vary in their implant doses, which are in general high and therefore also expected to impact the crystal quality. All samples began as purchased substrates polished by their manufacturer (details unknown) but while the HPHT samples were left as-received prior to implantation, Method 2 and 3 samples all underwent some level of CVD overgrowth designed in part to mitigate polishing-induced strain. 
%Near-surface strain depends both on the bulk crystal properties and on the surface treatment. 
CVD growth is known to be susceptible to the incorporation of growth defects such as crystal lattice dislocations that may result in localised strain \cite{Kehayias2019}, while the thermodynamically stable HPHT growth process can result in fewer strain-inducing extended growth defects. 
%However, our samples also vary in their surface finish (polished vs polished then overgrown) and the implant dosages as well as implant species vary. The HPHT samples, for example, are commercially polished and are implanted at high doses so may be expected to exhibit a higher level of strain variation than the $\delta$-doped samples, with their as-grown surfaces and much lower implant doses. 

\begin{figure}
\centering
\includegraphics[width=0.45\textwidth]{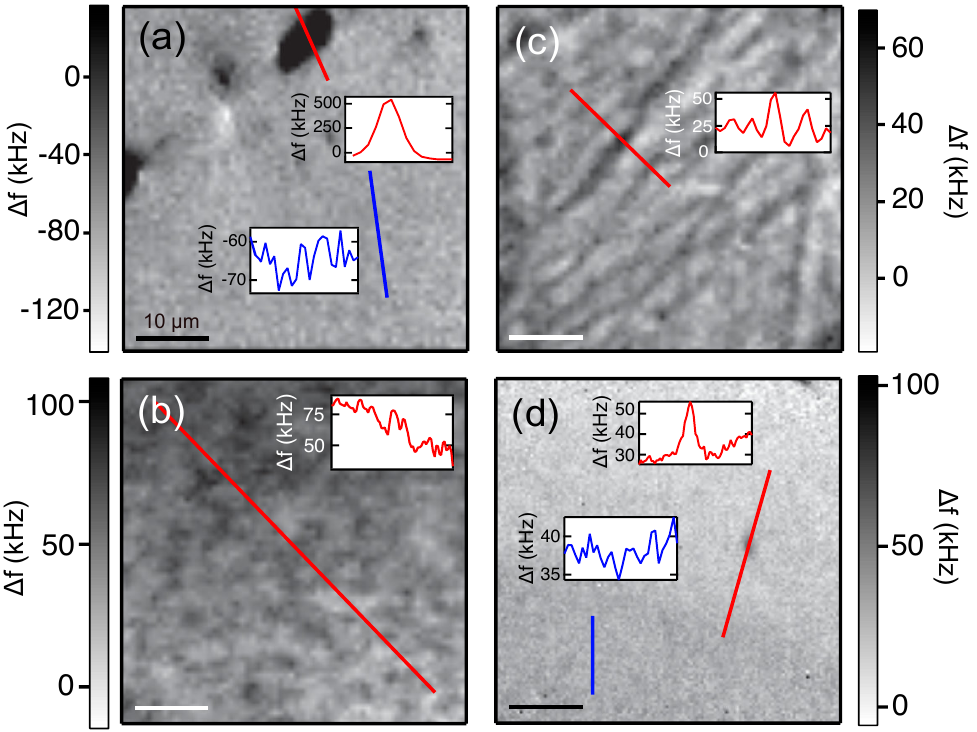}
\caption{\textbf{Representative strain maps from each method a)} Sample HPHT-4: HPHT sample; \textbf{b)} Sample N-5: N$^+$ implant; \textbf{c)} Sample CN-1: CN$^-$ implant sample; \textbf{d)} Sample $\delta$-2: $\delta$-doped sample. The images show the variation $\Delta f$ of the zero-field splitting constant $D$. The insets show linecuts taken along the red and blue solid lines.}
\label{fig: strain}
\end{figure}

To assess the strain homogeneity within our samples, we mapped the strain over the full 50~$\mu$m~$\times$~50~$\mu$m field of view using ODMR spectroscopy. ODMR maps focussing on the aligned $\ket{0} \rightarrow \ket{\pm 1}$ transitions under a bias field of 60~G were obtained for a variety of locations on a range of representative samples. From the two transition frequencies ($f_1$ and $f_2$) the zero-field splitting constant $D=\frac{f_1+f_2}{2}$ can be inferred as it is sensitive to strain but not magnetic fields \cite{Broadway2019}. Assuming temperature and electric fields are homogeneous across fields of view, $D$ maps thus tell us about the strain variations (in units of frequency) within samples. A series of representative strain maps for samples from each method is presented in Figure \ref{fig: strain}, where the colour scales denote the value of $\Delta f$, which is defined as the deviation from $D = 2870$~MHz (though the absolute value is arbitrary due to the possible co-presence of other fields which manifest as a constant shift across the field of view). 

In all four cases, strain variations over the scale of a few microns are present over the whole field of view. The amplitude of these variations ranges from a few kHz for the $\delta$-doped sample (close to the noise floor of the measurement) to 10-25~kHz in the others. In addition, stronger isolated features are sometimes visible, for example a 20~kHz spot in the $\delta$-doped sample, and 1~MHz streaks in the HPHT sample. We expect that the $\delta$-doped feature is an example of a CVD growth dislocation, while the HPHT features are polishing marks which are especially prevalent in the absence of any additional surface treatment. 

Background variation could originate in the crystal growth, polishing and implantation processes the samples underwent. The directionality of the pattern in Figure \ref{fig: strain}(b) suggests the substrate's polishing as the source, despite the subsequent 2~$\mu$m CVD overgrowth this sample underwent. This highlights the necessity of performing an additional process such as reactive ion etching (RIE), with or without subsequent overgrowth, to completely remove polishing damage \cite{Friel2009,Appel2016,Sangtawesin2019}. Sample CN-1 underwent an RIE process prior to overgrowth and Figure \ref{fig: strain}(c) shows that this was successful in removing obvious polishing features, although there is still a noticeable strain gradient and variation of up to 25~kHz. This background is unlikely to be due to the CN implantation alone as it is larger than samples subjected to higher doses and so is thought to be damage related to the RIE itself. This shows the importance of optimising the RIE process to best prepare the surface, following an Ar/Cl etch like the one conducted here with an additional oxygen etch \cite{Sangtawesin2019}.

The lowest level of background variation ($\sim 5$~kHz) was found in the $\delta$-doped sample (Figure \ref{fig: strain}(d)), which benefited from the lowest implantation dose as well as the thickest CVD overgrowth ($>10$~$\mu$m) on a polished (HPHT) substrate. HPHT substrates were chosen for these growths primarily to inherit their low dislocation densities in the CVD overgrowth \cite{Martineau2009}. An oxygen etch process designed to remove polishing damage was conducted prior to the CVD overgrowth and appears to have been successful in preventing the appearance of polishing features in the final grown layer \cite{Tallaire2004}. Large dislocation features are sparse and only of the same order as background variation in the other samples, indicating that although the CVD process is susceptible to the incorporation of crystal imperfections, this is a relatively small detraction compared to the polishing and implantation processes. This also shows that, with careful selection of growth conditions and underlying substrate, CVD growth is capable of producing mostly strain-free material.

The large polishing marks observed in the HPHT samples underline the importance of undertaking some form of post-polishing process. However they are in general quite sparse, meaning that quick pre-screening of such samples should be adequate to ensure an uncompromised imaging experiment. The background variation in these samples is of similar magnitude to that of Method 3 samples despite the polished surface finish, which is likely an indication of the high strain uniformity of the original crystal. 

It is reasonable to expect some degree of variation in the strain homogeneity between and even within samples, however we can still draw informative conclusions from these results. 
Even away from large features, typical background variation of tens of kHz is significant for high sensitivity magnetic imaging (corresponding to variations on the order of 1~$\mu$T). This means that the specifics of sample surface preparation, which the results of this section show is the single greatest contributor to strain homogeneity even when viewed alongside high ion implantation doses and variation in crystal growth, are especially important for future applications of widefield NV imaging. In contrast, the strain variations over a field of view are still sufficiently small in all cases to not limit the measured $T_2^*$ values presented in Section \ref{sec: results1}, hence $T_2^*$ is likely limited by magnetic noise and the effects of crystal strain will be less noticeable in measurements averaged over a field of view for this density of NV ensemble. 

\subsection{Temperature dependence}
\label{sec: temp}

Some applications of widefield NV imaging require the sample of interest to be cooled to cryogenic temperatures. For example, superconducting phenomena (vortices and transport currents) were recently imaged at 4~K \cite{Lillie2020}. Thus, it is useful to investigate the temperature dependence of the magnetic sensitivity of the NV layers and so we repeat the suite of measurements from Section \ref{sec: results1} as a function of temperature from 4-300~K for a series of representative samples.

\begin{figure}
\centering
\includegraphics[width=0.45\textwidth]{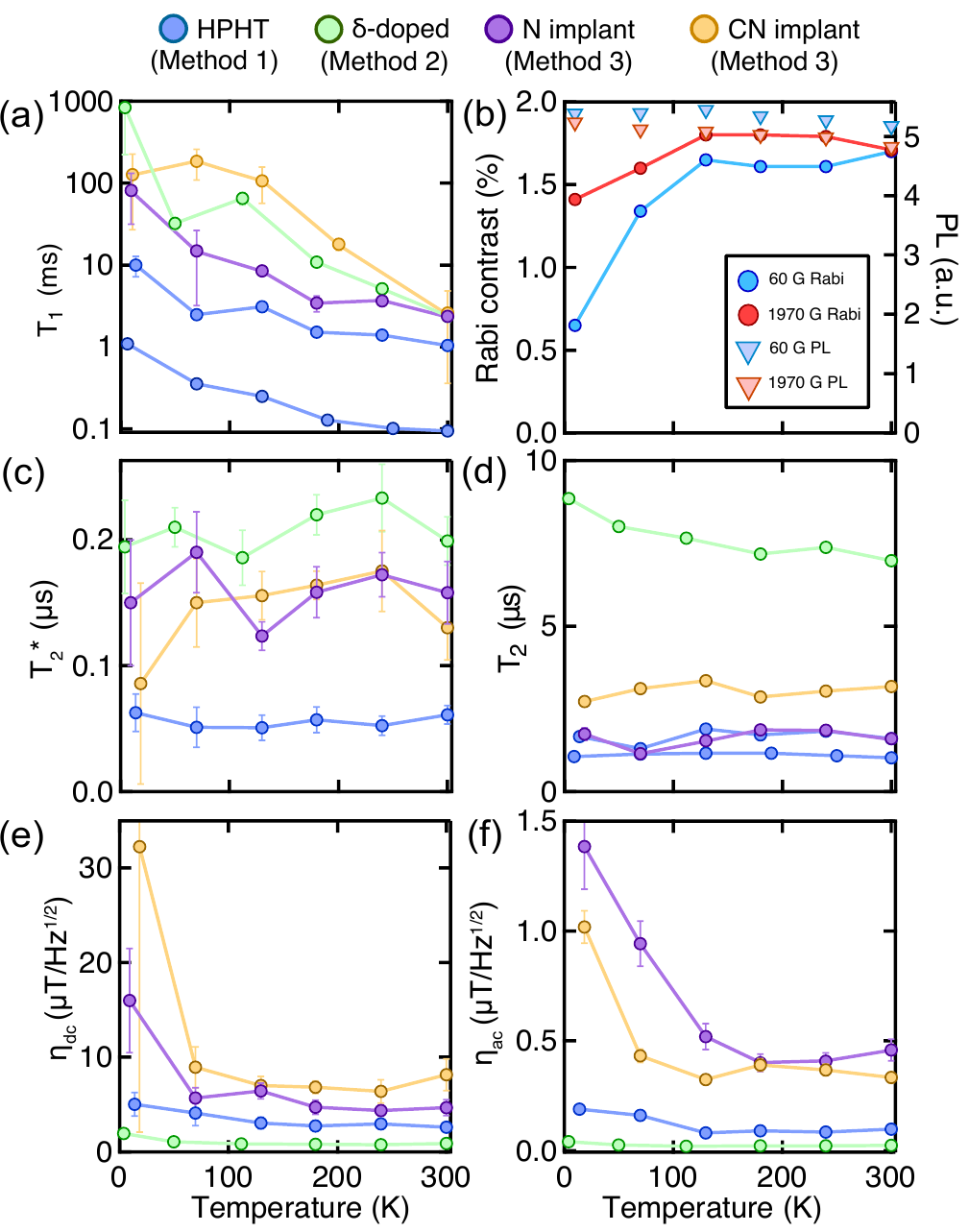}
\caption{\textbf{Spin property scaling with temperature:} \textbf{a)} $T_1$, plotted on a log scale.  \textbf{b):} Rabi oscillations taken at 4~K a field of 60~G (blue) and 1970~G (red) in sample CN-1 showing increased measurement contrast at higher fields. Points marked by triangles show PL measurements, showing reduced fluorescence at the higher field but not enough to explain the difference in contrast. \textbf{c)} and \textbf{d)} show $T_2^*$ and $T_2$ variation with temperature respectively. \textbf{e)} and \textbf{f)} show the DC and AC magnetic sensitivity scaling respectively. Data presented is for samples $\delta$-2 (green), N-2 (purple), CN-1 (yellow), and HPHT-1 (blue) in all comparative plots, with the addition of sample HPHT-3 as the bottom blue set in a) and d). }
\label{fig:cold}
\end{figure}
We first plot $T_1$ vs temperature in Figure \ref{fig:cold}(a).  At low temperatures, we see significant extension of $T_1$ for all samples. The largest extension (around two orders of magnitude) is seen for samples with long room temperature $T_1$, consistent with them being governed by phonon relaxation. However, even the samples with short room temperature $T_1$ show a significant extension (of around one order of magnitude for HPHT samples). This suggests that the dominant contribution to $T_1$ in this case depends on temperature as well, possibly through thermal activation of donor/acceptor defects, although further work is needed to fully understand this behaviour.

% For samples that display the most significant extension, the fitting exponent $k$ changes from $\sim$1 at room temperature to $\sim$0.5 at 4~K, confirming the change in dominant relaxation pathway from phonon-mediated to cross relaxation \cite{Bauch2019}. 

In Figure \ref{fig:cold}(b) we plot Rabi contrast (Sample CN-1) against temperature and see a surprising reduction at 60~G below 100~K. This trend is consistently observed but is most marked for N implants (a factor of 3), while $\delta$-doped samples saw a lesser reduction (factor of 2), though it is unclear from the small sample size whether this difference is significant. This reduction is largely recovered  at a higher field of 1970~G. This phenomenon has previously been observed in single NV centres \cite{privcom} and its origin is currently being investigated. The PL (also plotted in Figure \ref{fig:cold}(b)) does not vary significantly with temperature and although there is a minor reduction in PL at the higher field due to increased quenching of off-axis NVs \cite{Tetienne2012}, this is not enough to explain the differences in measured contrast. No other spin property was observed to change between the two fields apart from a slight extension of $T_1$ at 1970~G.

Figure \ref{fig:cold}(c) shows the $T_2^*$ variation with temperature, which is very small for all samples, indicating that temperature-independent magnetic noise dominates this quantity at all temperatures. Hahn echo $T_2$ (Figure \ref{fig:cold}(d)) is also broadly constant with temperature, with the exception of a $\approx 25$\% extension for the $\delta$-doped sample with long room temperature $T_2$. Decoupled, saturated $T_2$ has previously been shown to have a phonon-coupled component for dilute spin baths in bulk and near-surface regimes \cite{Romach2015,Bar-Gill2013}. Though we only consider Hahn echo AC sensing here for simplicity, the slight extension at low temperatures for the sample with the least dense spin bath suggests a similar phonon component to $T_2$ which would be more obvious for higher order decoupling sequences. 

In Figure \ref{fig:cold}(e) and (f), the measured quantities are propagated to magnetic sensitivities to DC and AC fields respectively at 60~G. There is little variation in sensitivity with temperature except for the deterioration at low temperature due to reduction in spin contrast $\cal C$. The absolute values are higher than in Section \ref{sec: results1} due to the lower laser power and lower collection efficiency on the cryostat system but the ordering between methods remains unchanged, meaning the conclusions drawn in the previous sections remain valid. This means that although this measurement regime does require additional attention due to non trivial factors such as the contrast loss (resulting in reduced sensitivity below 100~K), the suitability of samples themselves is similar to the room temperature case and so the findings of this study are equally applicable in the design of low temperature experiments. 

\section{Discussion} \label{sec:discussion}

We now discuss the pros and cons of each method and routes for improvement. 

Method 1 is by far the most cost effective and requires only a simple implantation procedure as the HPHT substrates used are commercially available at relatively low cost. Carbon irradiation provides the required depth profile tunability and unnecessary damage to the crystal can be avoided by tuning the process's vacancy creation to the existing N density. Relying on commercial substrates does mean, however, that this method lacks control over the N density. In addition, the exact concentration is not known ahead of time as it is only loosely specified by manufacturers, and can even vary within individual diamonds as they frequently contain different growth sectors that incorporate defects at different rates. This variability between (and within) samples is present in our results as a greater spread in sensitivity within this method than the others, though further optimisation of the procedure is expected to reduce this. The N density in these samples is usually quite high ($\sim$100~ppm), resulting in shorter coherence times and this has typically been seen as a detraction, resulting in HPHT diamonds seeing little use for widefield NV imaging to date. However, our results show that, following implantation, a bright ensemble providing enough fluorescence to offset the short coherence times is produced. Our data shows that the best (most optimised) HPHT samples were among the highest sensitivity samples studied.

No pre-measurement of nitrogen concentration was undertaken for this work and so there is obvious room for further optimisation in the future. Additional studies into the merit of implanting with lighter ions such as helium could also improve the outlook, producing more even damage to the lattice at the cost of introducing defects related to the implant species.

Like Method 1, in situ $\delta$-doping (Method 2) has the advantage of naturally incorporating nitrogen during growth, allowing the creation of high-yield ensembles following tuned vacancy production at a minimal damage cost. Being a custom process, $\delta$-doping allows control over N density and the depth profile of the doped layer. These benefits are, however, difficult to realise and reproduce due to the complexity of the process and it is in general the most cost intensive of the three methods. CVD growth also carries the detraction of possible incorporation of unwanted point defects such as the NVH complex that limits the NV conversion ratio \cite{Stacey2012a}, as well as crystal dislocation defects that may introduce additional strain \cite{Kehayias2019}. The results of Section \ref{section: strain}, however, show that material with a high level of strain uniformity can be produced using growth conditions designed to promote slow growth and careful selection and preparation of the growth substrate. Additionally, it appears that the implant doses useful for creating the high density ensembles that this study focusses on, and in particular the surface preparation of the diamond, play more prominent roles in decreasing strain homogeneity than the method of diamond growth. 

Samples grown in the slightly lower density regime (Samples $\delta$-3 and $\delta$-4, with an order of magnitude lower NV density than other Method 2 samples) were observed to be susceptible to uneven nitrogen incorporation. These low density samples exhibited fluorescence striations expected to have arisen as a result of step-bunching growth patterns that see nitrogen defects preferentially incorporate at step-edges. This unevenness is undesirable for widefield imaging, especially in a modality that may seek to image via fluorescence quenching, but would be less relevant in confocal microscopy. This indicates that while in principle $\delta$-doping is highly tunable, the complexity of the process requires careful refinement to realise this. This behaviour appears to be overridden at higher nitrogen densities as the other samples exhibited highly even fluorescence, though the as-grown surface could still contain unwanted features such as growth hillocks that are undesirable for imaging experiments. 

The complexity of the CVD growth process also means that there exists a lot of room for improvement within this method, despite it already producing the highest magnetic sensitivities. In addition to the improvements identified for Method 1, we note that further optimisation of (and finer control over) other key growth parameters such as gas composition and growth temperature could see higher quality crystals with fewer unwanted defects. Confining nitrogen to a well-defined layer during growth also carries with it the advantage of being able to explore the merits of introducing vacancies via electron irradiation~\cite{Kim2012,McLellan2016} rather than the implantation of more massive particles. This possible improvement is incompatible with Method 1 as electrons cannot be confined in depth in the same way as carbon ions.

Method 3 has the advantages of giving a predictable N density and depth profile that can be easily tuned. However, this comes at the considerable cost of excess vacancy creation (about two orders of magnitude at the energies considered, as shown in Figure \ref{fig:schematic}(c)) that cannot be tuned. This manifests itself in our results as reduced coherence values relative to nitrogen concentration compared to samples that have N and V concentrations roughly matched, ultimately resulting in inferior magnetic sensitivity. This limitation will be inherent to any procedure that relies on incorporating nitrogen into the crystal via implantation unless new techniques allow better vacancy removal during annealing (discussed below). This method also requires a high purity CVD substrate which are generally more costly than HPHT substrates.

Comparing the two species used to investigate this method, our results show that implanting with the larger CN$^-$ ion results in more damage to the crystal per ion implanted, compromising both AC and DC sensitivity. Much of the AC sensitivity can be recovered, however, by using longer dynamical decoupling sequences, indicating the effectiveness of this approach in protecting spin coherence against interactions with unwanted paramagnetic defects.

Finally, method-agnostic improvements were beyond the scope of this work but may include dynamic annealing during the implantation stage to better repair crystal damage, and surface treatments to limit surface noise~\cite{Kucsko2018,Sangtawesin2019}. Optimised RIE may be useful in mitigating strain caused by polishing damage, resulting in a more homogeneous zero field splitting across a field of view to be imaged and possibly improving NV coherence by reducing surface magnetic noise \cite{DeOliveira2015}. Fermi level engineering via n-type doping (phosphorus, sulphur) has been shown to promote the NV$^-$ charge state by charging vacancies and other defects which then can be annealled out to improve the NV yield \cite{Luhmann, Herbschleb}. It is unclear whether this will have a significant effect in the higher nitrogen concentration regime wherein the majority of samples measured in this study belong, so further work is required in this area.  

\section{Conclusion} \label{sec:conclusion}

Our results show that while $\delta$-doping during CVD growth offers the best overall sensitivity across a range of temperatures from cryogenic to room temperature, implantation of nitrogen-rich HPHT substrates is highly competitive, offering superior sensitivities to the much more expensive and currently ubiquitous approach of implanting electronic-grade CVD substrates with nitrogen. Though the HPHT approach has a significant drawback in its lack of reproducibility and precise control over NV density compared to current alternatives, its low cost and untapped potential for further improvement by refinement of the implantation process leaves it as an appealing option for future research.

Even though the CVD growth for the $\delta$-doped samples presented here is not expected to be state-of-the-art, the best sensitivities were achieved by these samples. This indicates that, if the cost requirements are not too strenuous, $\delta$-doping is an achievable method for the creation of high quality NV ensembles of varying density.

The common method of nitrogen implantation resulted in the lowest magnetic sensitivity of the three methods. This is due to the large amounts of damage sustained by the crystal during the high dose implantation procedures required to incorporate the high levels of nitrogen desired for ensemble NV sensing. Unlike the other two methods, implanting nitrogen leaves no scope to tune vacancy production to nitrogen density and this looms as the main limitation of Method 3.  

It is clear from our results that the high nitrogen content of the HPHT samples gives sufficiently high PL following C$^-$ implantation to offset their short coherence times. Additionally, although it offers considerably less flexibility than the other two methods, high PL HPHT samples are particularly advantageous for measurements with very low duty cycles (e.g. $T_1$ measurements) or very low laser power (e.g. to minimise sample heating \cite{Lillie2020}). Indeed, in these situations noise sources other than shot noise, such as dark currents in an sCMOS camera, may be dominant if the PL level is too low, deteriorating the sensitivity. The flexibility of $\delta$-doping makes it suitable for the widest range of applications, with the lower density samples excelling in applications that allow long measurement times to reach the shot noise limit. Method 3 is shown to be suboptimal in both absolute sensitivity terms and in practicality from a cost point of view in this work, however we note that further advances in annealing procedures to repair crystal lattice damage sustained during implantation could entirely address the losses in sensitivity. The NV layer formation methods themselves appear to have relatively little impact on the strain homogeneity of the final crystal compared to the diamond surface preparation. The optimisation of these processes for enhancing strain homogeneity and assessing the resulting improvement on NV ensemble quality for widefield imaging is left for future work.

\section*{Acknowledgements}

We acknowledge support from the Australian Research Council (ARC) through grants DE170100129, DE190100336, LP160101515, CE170100012, LE180100037 and DP190101506. We also acknowledge the AFAiiR node of the NCRIS Heavy Ion Capability for access to ion-implantation facilities. This work was performed in part at the Melbourne Centre for Nanofabrication (MCN) in the Victorian Node of the Australian National Fabrication Facility (ANFF). A.J.H. and D.A.B. are supported by an Australian Government Research Training Program Scholarship. T.T. acknowledges the support of JSPS KAKENHI (No. 20H02187 and 19H02617), JST CREST (JPMJCR1773) and MEXT Q-LEAP (JPMXS0118068379).

\appendix

\section{Full room temperature results}

\label{app:results}
In Table \ref{Table:results} we list the full data plotted in Figures \ref{Fig: coherence}, \ref{fig: sensitivity}, and \ref{fig: cpmg} for comparison with the parameters in Table \ref{Table:samples}. 
\begin{table*}[hbt!]
\begin{tabular}{|c|c|c|c|c|c|c|c|c|}
\hline
Sample & $\cal C$ (\%) & $T_2^*$ ($\mu$s) & $T_2^{\text{HE}}$ ($\mu$s) & $T_2^{\text{CPMG}}$ ($\mu$s) & $T_1$ (ms) & $\eta_{\rm dc}$ & $\eta_{\rm ac}$ & $\eta_{\rm CPMG}$ \\
name &  &  &  &  &  & ($\mu$T Hz$^{-1/2}$) & (nT Hz$^{-1/2}$) & (nT Hz$^{-1/2}$) \\ \hline
HPHT-1 & 3.0~$\pm~ 0.1$ & 0.088~$\pm~ 0.020$ & 1.04~$\pm~ 0.03$ & 60.2~$\pm~ 0.7$ & 0.35~$\pm~ 0.01$ & 1.10~$\pm~ 0.31$ & 99.5~$\pm~ 17.1$ & 5.09~$\pm~ 0.86$ \\
HPHT-2 & 2.2~$\pm~ 0.1$ & 0.024~$\pm~ 0.080$ & 0.94~$\pm~0.01$ & 63.5~$\pm~ 0.2$ & 0.29~$\pm~ 0.01$ & 5.83~$\pm~ 1.32$ & 155.7 ~$\pm~ 23.8$ & 7.10~$\pm~ 1.05$ \\
HPHT-3 & 1.5~$\pm~ 0.1$ & 0.025~$\pm~ 0.008$ & 0.67~$\pm~0.02$ & 4.5*~$\pm~ 0.01 $ & 0.02~$\pm~ 0.01$ & 4.42~$\pm~ 1.43$ & 174.0~$\pm~ 12.9$ & 32.36*~$\pm~ 2.15$ \\
HPHT-4 & 1.6~$\pm~ 0.1$ & 0.042~$\pm~ 0.013$ & 0.89~$\pm~0.01$ & 9.2*~$\pm~ 0.07$ & 0.10~$\pm~ 0.01$ & 4.71~$\pm~ 1.69$ & 233.6~$\pm~ 43.5$ & 32.88*~$\pm~ 6.12$ \\
HPHT-5 & 1.5~$\pm~ 0.1$ & 0.029~$\pm~ 0.010$ & 0.49~$\pm~0.05$ & 3.5*~$\pm~ 0.01$ & 0.11~$\pm~ 0.01$ & 3.77~$\pm~ 1.46$ & 234.7~$\pm~50.9$ & 39.07*~$\pm~ 7.35$ \\
HPHT-6 & 1.8~$\pm~ 0.1$ & 0.016~$\pm~ 0.007$ & 0.83~$\pm~0.01$ & 58.2~$\pm~ 23.80$ & 0.06~$\pm~ 0.01$ & 6.05~$\pm~ 2.93$ & 120.5~$\pm~ 16.5$ & 5.09~$\pm~ 2.19$ \\
HPHT-7 & 3.0~$\pm~ 0.1$ & 0.012~$\pm~ 0.009$ & 1.17~$\pm~0.05$ & 50.9~$\pm~ 4.00$ & 0.14~$\pm~ 0.01$ & 0.49 ~$\pm~ 0.08$ & 54.2~$\pm~ 8.3$ & 3.41~$\pm~ 0.57$ \\
$\delta$-1 & 4.0~$\pm~ 0.1$ & 0.347~$\pm~ 0.058$ & 8.19~$\pm~0.16$ & 285.3~$\pm~ 0.17$ & 1.88~$\pm~ 0.10$ & 0.52 ~$\pm~ 0.10$ & 32.1~$\pm~ 3.3$ & 4.11 ~$\pm~ 0.42$ \\
$\delta$-2 & 3.8~$\pm~ 0.1$ & 0.221~$\pm~ 0.098$ & 7.49~$\pm~0.12$ & 301.8~$\pm~ 10.00$ & 1.54~$\pm~ 0.07$ & 0.64~$\pm~ 0.30$ & 27.1~$\pm~ 3.8$ & 3.16~$\pm~ 0.45$ \\
$\delta$-3 & 1.2~$\pm~ 0.1$ & 1.238~$\pm~ 0.173$ & 152.53~$\pm~10.78$ & 1723.5~$\pm~ 500.00$ & 3.04~$\pm~ 0.09$ & 2.22~$\pm~ 0.34$ & 81.7~$\pm~ 10.7$ & 23.86~$\pm~ 7.06$ \\
$\delta$-4 & 3.1~$\pm~ 0.1$ & 0.742~$\pm~ 0.147$ & 71.28~$\pm~6.95$ & 700.0~$\pm~ 200.00$ & 3.00~$\pm~ 0.10$ & 1.04~$\pm~ 0.22$ & 35.5~$\pm~ 4.6$ & 10.91~$\pm~ 3.25$ \\
$\delta$-5 & 4.0~$\pm~ 0.1$ & 1.014~$\pm~ 0.111$ & 8.79~$\pm~0.15$ & 332.1~$\pm~ 50.00$ & 1.72~$\pm~ 0.04$ & 0.18~$\pm~ 0.03$ & 29.1~$\pm~ 4.0$ & 3.62~$\pm~ 0.74$ \\
N-1 & 2.4~$\pm~ 0.1$ & 0.071~$\pm~ 0.009$ & 2.46~$\pm~0.12$ & 51.6~$\pm~ 5.60$ & 0.50~$\pm~ 0.02$ & 5.23~$\pm~ 1.08$ & 176.2~$\pm~ 29.5$ & 21.36~$\pm~ 4.13$ \\
N-2 & 2.9~$\pm~ 0.1$ & 0.157~$\pm~ 0.009$ & 2.27~$\pm~0.05$ & 30.0*~$\pm~ 0.90$ & 0.61~$\pm~ 0.02$ & 1.56~$\pm~ 0.27$ & 124.8~$\pm~ 20.3$ & 19.28*~$\pm~ 3.15$ \\
N-3 & 2.3~$\pm~ 0.1$ & 0.109~$\pm~ 0.011$ & 4.20~$\pm~0.01$ & 64.9~$\pm~ 15.00$ & 1.05~$\pm~ 0.05$ & 2.78~$\pm~ 0.49$ & 92.9~$\pm~ 13.0$ & 15.54~$\pm~ 4.20$ \\
N-4 & 2.8~$\pm~ 0.1$ & 0.060~$\pm~ 0.007$ & 1.36~$\pm~0.01$ & 41.7~$\pm~ 4.20$ & 0.32~$\pm~ 0.02$ & 3.96~$\pm~ 0.82$ & 193.0~$\pm~ 32.2$ & 15.53~$\pm~ 3.02$ \\
N-5 & 3.1~$\pm~ 0.1$ & 0.130~$\pm~ 0.012$ & 2.67~$\pm~0.09$ & 61.0~$\pm~ 4.70$ & 0.54~$\pm~ 0.03$ & 2.05~$\pm~ 0.34$ & 117.9~$\pm~ 16.8$ & 13.98~$\pm~ 2.21$ \\
CN-1 & 2.9~$\pm~ 0.1$ & 0.092~$\pm~ 0.025$ & 2.69~$\pm~0.11$ & N-1.8~$\pm~ 4.80$ & 0.77~$\pm~ 0.05$ & 4.51~$\pm~ 1.52$ & 181.1~$\pm~ 35.7$ & 10.60~$\pm~ 2.06$ \\
CN-2 & 3.3~$\pm~ 0.1$ & 0.080~$\pm~ 0.033$ & 2.26~$\pm~0.06$ & 112.1~$\pm~ 6.50$ & 0.66~$\pm~ 0.04$ & 5.03~$\pm~ 1.52$ & 204.1 ~$\pm~ 35.2$ & 15.14~$\pm~ 2.73$ \\
CN-3 & 3.6~$\pm~ 0.1$ & 0.123~$\pm~ 0.038$ & 3.75~$\pm~0.09$ & 129.7~$\pm~ 6.00$ & 0.64~$\pm~ 0.04$ & 3.33~$\pm~1.31$ & 124.2~$\pm~ 23.2$ & 13.09~$\pm~ 2.77$ \\ \hline
\end{tabular}
\caption{\textbf{Full spin properties:} Summary of room temperature data presented in Section \ref{sec:results}. All measurements as defined earlier, with $T_2^{\text{CPMG}}$ being the CPMG-1024 $T_2$ value (or highest $N$ CPMG sequence for samples that could not be extended this far, marked with an asterisk).}	
\label{Table:results}	
\end{table*}

\section{Data collection}
\label{app: measurements}

In this appendix we provide greater detail on the measurements undertaken as introduced in the main text. 

\subsubsection{Photoluminescence (PL)}

The PL rate, $\cal R$, was measured by averaging over an image taken under constant laser illumination over a 30~ms camera exposure and a 50~$\mu$m~$\times$~50~$\mu$m area. The per-pixel average obtained was converted into a per $\mu$m$^2$ per second value as a figure of merit. $\cal R$ was then converted into an approximate areal NV density by comparing the fluorescence of a representative set of samples on a confocal microscope (recorded by an avalanche photo diode) with that from a single NV (from a sample separate to this study). Multiple samples were compared to confirm an appropriate multiplicative factor to convert the widefield PL measurements to this density. Minor background fluorescence from the HPHT substrate present in the $\delta$-doped samples was subtracted. 
%but could be trivially subtracted as the offset from zero of the Gaussian PL profile as viewed on the full field of view with the laser focussed on the sensing layer. 

\subsubsection{Rabi contrast}

${\cal C}$ was calculated by driving the NV ensemble at a high enough microwave power for the $T_{1\rho}$ decay to be negligible over the first Rabi oscillation and fitting, with $\cal C$ defined as the peak-to-peak amplitude. This is a measure of the maximum attainable contrast, an upper bound for the actual contrast available for a measurement optimised at $T_2$ or $T_2^*$. 

As with $\cal R$, the impact of background fluorescence artificially decreasing $\cal C$ was removed as a less fluorescent substrate could be used in future without affecting NV properties.  
%The measured contrast for the lowest density CVD samples suffered from residual fluorescence from the HPHT substrate. Other impact from the substrate is expected to be minimal due to the electronic-grade spacing layer onto which the delta doped film was grown, but our widefield setup does not allow for the spatial filtering in $z$ found in confocal microscopes so this background could not be removed. As this is not a fundamental limitation of the method and could be easily remedied by using a confocal microscope or a less fluorescent substrate, the measured contrasts for the CVD samples were adapted by removing the background as above. 

The lowest density CVD samples also exhibited PL striations arising step bunching growth behaviour. This meant the PL was variable across the field of view used and is also expected to reduce the contrast when averaging over the full region. This effect was not removed as it was observed to be a limitation inherent to growing at low nitrogen concentrations under our growth conditions. 

\subsubsection{Free induction decay time}

$T_2^*$ was measured using the Ramsey pulse sequence with a level of detuning of the microwave drive from the hyperfine resonances adjusted to each sample to ensure an appropriate frequency of beating to allow accurate extraction from the resulting fit. The signal was fit to a sum of cosines with frequencies given by the detuning from the respective hyperfine resonances modulated by an exponential envelope with decay constant $T_2^*$, $e^{-\left(\tau / T_2^* \right)} \sum_{i=1}^{N} \cos\left(\omega_i\tau + \phi_i\right)$, where the $\omega_i$ are detunings from the individual hyperfine resonances, $\phi_i$ are phases, and $N$ is the number of hyperfine resonances, equalling 3 and 2 for $^{14}$N and $^{15}$N respectively. Care was taken to ensure the measurement was not compromised by magnetic field gradients across the field of view so that the measured decay envelope could accurately be attributed to the free induction decay of the ensemble alone. 

\subsubsection{Spin relaxation time}

$T_1$ was measured by initialising the NV ensemble and reading out the state after set periods of dark evolution. The signal was normalised by subtracting the PL obtained by a sequence identical but for the application of a $\pi$ pulse prior to readout. This is a standard procedure in our measurements but is particularly important in the measurement of $T_1$ due to the wide range of wait times involved that result in a variable laser duty cycle and thus changing charge state dynamics over the course of a single sweep. The resulting decay was fit to a stretched exponential $e^{-\left(\tau/T_1\right)^k}$ with $k$ ranging between 0.5 and 1 at room temperature.

\subsubsection{Spin coherence time}

The sensitivity of the samples to AC fields was measured using a Hahn spin echo sequence at a field of 475~G chosen to ensure that, for the samples containing a natural abundance of $^{13}$C, the resulting revivals were closely spaced enough to still allow us to accurately extract $T_2$ from the decay curve. The envelope of the resulting curves were fit to an exponential decay $e^{-\left(\tau/T_2\right)^k}$ and the Hahn echo spin coherence time $T_2$ extracted. $k$ varied between 0.5 and 1.5. 

\subsubsection{Extended dynamical decoupling}

The behaviour of the samples with respect to extended decoupling was assessed using CPMG sequences of increasing length until either $T_2$ saturation or the total microwave duty cycle exceeded the sample's $T_{1\rho}$ (and hence no further meaningful extension could be achieved). These measurements were carried out at 475~G so as to suppress the impact of any $^{13}$C revivals on our ability to fit the resulting curve and reliably extract the decay constant. Fluorescence for all measurements of a sample were normalised to the contrast obtained for the shortest decoupling sequence ($\approx \cal C$) and fit to an exponential decay of form $ e^{-\left( \tau /T_2\right)^k}$, modulated by $^{13}$C revival structure fit as Lorentzian dips where appropriate. No strong, consistent trend in the variation of $k$ with $N$ was observed across samples. 
\subsubsection{Strain mapping}
%Under a bias field of 60~G the ODMR spectrum around the expected on-axis transitions ($\ket{0}\rightarrow\ket{\pm 1}$) was taken using a low power microwave drive to achieve $T_2^*$-limited linewidths. 
To obtain maximum fidelity, the ODMR spectra used to obtain crystal strain maps utilised low power microwave driving to achieve $T_2^*$-limited linewidths. The resulting data was fit to Lorentzian lineshapes (accounting for hyperfine structure where appropriate) and each resonant frequency extracted for each pixel. Pixel size was set to be comparable to the optical diffraction limit $\approx$300~nm. Images were saved periodically until the level of pixel-to-pixel noise reached a plateau, indicating that photon shot noise was no longer the dominant source of noise and any remaining variation could be attributed to strain inhomogeneity. 
%All effects from magnetic fields were normalised out by summing the resonant frequencies leaving only contributions due to stress and strain which shift all resonances in the same direction. 
%Full vector imaging of stress features cannot be achieved with this technique as we only address one set of NV orientations, however we can obtain a measure of the magnitude of characteristic features, as well as the limiting noise floor once shot noise becomes less significant than the pixel-to-pixel strain inhomogeneity (if it does).
%\subsubsection{Variation in properties with temperature}
%With the exception of ODMR mapping and extended dynamical decoupling, the above measurements were repeated over a range of temperatures on a setup capable of reaching temperatures down to 4~K for representative samples of each class. As well as being conducted at a field of 60~G for best comparison to the room temperature measurements, these measurements were repeated at a field of $\sim$~2000~G at each temperature step to investigate the previously observed phenomenon of Rabi contrast enhancement at this field at low temperatures. 

\subsubsection{Low temperature measurements}
The cryogenic widefield setup is functionally identical to that used to take room temperature measurements and so measurements taken on it proceeded as described above. The need to use a low laser power (to avoid sample heating) and the lower collection efficiently as well as the loss of measurement contrast at low temperatures increased the measurement uncertainty, particularly for free induction decay measurements on less fluorescent samples (ie CN$^-$ implants). Similarly, there is a large uncertainty in the measurement of long ($\gg 10$~ms) $T_1$ decays as the readout duty cycle becomes insignificant next to the wait times and so camera dark currents (and their associated noise) begin to account for a large proportion of the total signal. Nevertheless, $T_1$ extension with decreasing temperature is clear, which is the main observation to make from this section. 

Although the base temperature of the cryostat is 4~K, the conditions used to access the full range of temperatures without otherwise altering the experiment meant that the cooling power was not always sufficient to maintain this temperature under the influence of laser heating. The temperature was measured after every sweep of one camera exposure per data point and then averaged for plotting. The temperature throughout an experiment was usually mostly constant as the laser duty cycle does not change much from point to point. The laser heating effect was most significant for the Sample CN-1 $T_2^*$ measurement as this was the closest to a CW laser duty cycle and undertaken under conditions of least cooling power. A minimal number of sweeps were used to maintain a low temperature for this measurement, which contributed to its high uncertainty. 

Temperatures above the base temperature were accessed using a heater placed below the sample and fluctuations within these measurements were minimal. Room temperature measurements conducted on this system occasionally differed slightly from the original set due to differences in ensemble initialisation and microwave driving power and so data is directly comparable within the context of individual sections only. 

\section{Further sample details}
\label{sampledetails}

The samples and their creation methods have been described in the main text, but here we note additional features of some of the samples used. Samples N-1, N-5, HPHT-7 and HPHT-2 had an additional coating on the NV side (80~nm Al and 75 nm of A$_2$O$_3$) though this is not expected to affect the measurements presented here. Sample N-4 was imaged face-up as it was mounted as part of a device for separate work, though this is also not thought to impact the following measurements as there is no thick substrate present to hinder fluorescence collection.

HPHT substrates used for the production of Method 2 samples first underwent a 3~hour oxygen etch at 90~Torr~/~5200~W (15~sccm O$_2$, 478~sccm H$_2$) before a two-stage N-free CVD overgrowth of 2~hours 85~Torr~/~5200~W and 12~hours 85~Torr~/~4500~W, each with gas flows of 20~sccm CH$_4$ and 450~sccm H$_2$. The sample temperature was measured to be approximately 850~$\degree$C during growth.

Sample $\delta$-5 was polished and subjected to an RIE treatment prior to $\delta$-doping. The surface preparation has previously been shown to make no significant difference to the coherence properties of dense ensembles \cite{Tetienne2018a}, so the effects of these processes are expected to be contained within the bounds of the discussion in Section \ref{section: strain}.

\bibliographystyle{naturemag}
\bibliography{bib}

\begin{thebibliography}{10}
\expandafter\ifx\csname url\endcsname\relax
  \def\url#1{\texttt{#1}}\fi
\expandafter\ifx\csname urlprefix\endcsname\relax\def\urlprefix{URL }\fi
\providecommand{\bibinfo}[2]{#2}
\providecommand{\eprint}[2][]{\url{#2}}

\bibitem{Doherty2013}
\bibinfo{author}{Doherty, M.~W.} \emph{et~al.}
\newblock \bibinfo{title}{{The nitrogen-vacancy colour centre in diamond}}.
\newblock \emph{\bibinfo{journal}{Physics Reports}}
  \textbf{\bibinfo{volume}{528}}, \bibinfo{pages}{1--45}
  (\bibinfo{year}{2013}).

\bibitem{Rondin2014}
\bibinfo{author}{Rondin, L.} \emph{et~al.}
\newblock \bibinfo{title}{{Magnetometry with nitrogen-vacancy defects in
  diamond}}.
\newblock \emph{\bibinfo{journal}{Rep. Prog. Phys.}}
  \textbf{\bibinfo{volume}{77}}, \bibinfo{pages}{56503} (\bibinfo{year}{2014}).

\bibitem{Steinert2010}
\bibinfo{author}{Steinert, S.} \emph{et~al.}
\newblock \bibinfo{title}{{High sensitivity magnetic imaging using an array of
  spins in diamond}}.
\newblock \emph{\bibinfo{journal}{Rev. Sci. Instrum.}}
  \textbf{\bibinfo{volume}{81}}, \bibinfo{pages}{043705}
  (\bibinfo{year}{2010}).

\bibitem{Pham2011}
\bibinfo{author}{Pham, L.~M.} \emph{et~al.}
\newblock \bibinfo{title}{{Magnetic field imaging with nitrogen-vacancy
  ensembles}}.
\newblock \emph{\bibinfo{journal}{New J. Phys.}} \textbf{\bibinfo{volume}{13}},
  \bibinfo{pages}{045021} (\bibinfo{year}{2011}).

\bibitem{Chipaux2015}
\bibinfo{author}{Chipaux, M.} \emph{et~al.}
\newblock \bibinfo{title}{{Magnetic imaging with an ensemble of NV centers in
  diamond}}.
\newblock \emph{\bibinfo{journal}{Eur. Phys. J. D}}
  \textbf{\bibinfo{volume}{69}}, \bibinfo{pages}{166} (\bibinfo{year}{2015}).

\bibitem{Simpson2016}
\bibinfo{author}{Simpson, D.~A.} \emph{et~al.}
\newblock \bibinfo{title}{{Magneto-optical imaging of thin magnetic films using
  spins in diamond}}.
\newblock \emph{\bibinfo{journal}{Sci. Rep.}} \textbf{\bibinfo{volume}{6}},
  \bibinfo{pages}{22797} (\bibinfo{year}{2016}).

\bibitem{Glenn2015}
\bibinfo{author}{Glenn, D.~R.} \emph{et~al.}
\newblock \bibinfo{title}{{Single-cell magnetic imaging using a quantum diamond
  microscope}}.
\newblock \emph{\bibinfo{journal}{Nat. Methods}} \textbf{\bibinfo{volume}{12}},
  \bibinfo{pages}{736} (\bibinfo{year}{2015}).

\bibitem{Tetienne2017}
\bibinfo{author}{Tetienne, J.-P.} \emph{et~al.}
\newblock \bibinfo{title}{{Quantum imaging of current flow in graphene}}.
\newblock \emph{\bibinfo{journal}{Sci. Adv.}} \textbf{\bibinfo{volume}{3}},
  \bibinfo{pages}{e1602429} (\bibinfo{year}{2017}).

\bibitem{Casola2018}
\bibinfo{author}{Casola, F.}, \bibinfo{author}{{Van Der Sar}, T.} \&
  \bibinfo{author}{Yacoby, A.}
\newblock \bibinfo{title}{{Probing condensed matter physics with magnetometry
  based on nitrogen-vacancy centres in diamond}}.
\newblock \emph{\bibinfo{journal}{Nature Reviews Materials}}
  \textbf{\bibinfo{volume}{3}}, \bibinfo{pages}{17088} (\bibinfo{year}{2018}).

\bibitem{Broadway2018c}
\bibinfo{author}{Broadway, D.~A.} \emph{et~al.}
\newblock \bibinfo{title}{{Spatial mapping of band bending in semiconductor
  devices using in situ quantum sensors}}.
\newblock \emph{\bibinfo{journal}{Nature Electronics}}
  \textbf{\bibinfo{volume}{1}}, \bibinfo{pages}{502--507}
  (\bibinfo{year}{2018}).

\bibitem{Fu2014}
\bibinfo{author}{Fu, R.~R.} \emph{et~al.}
\newblock \bibinfo{title}{{Solar nebula magnetic fields recorded in the
  Semarkona meteorite}}.
\newblock \emph{\bibinfo{journal}{Science}} \textbf{\bibinfo{volume}{346}},
  \bibinfo{pages}{1089--1092} (\bibinfo{year}{2014}).

\bibitem{Glenn2017}
\bibinfo{author}{Glenn, D.~R.} \emph{et~al.}
\newblock \bibinfo{title}{Micrometer-scale magnetic imaging of geological
  samples using a quantum diamond microscope}.
\newblock \emph{\bibinfo{journal}{Geochemistry, Geophysics, Geosystems}}
  \textbf{\bibinfo{volume}{18}}, \bibinfo{pages}{3254--3267}
  (\bibinfo{year}{2017}).

\bibitem{LeSage2013}
\bibinfo{author}{{Le Sage}, D.} \emph{et~al.}
\newblock \bibinfo{title}{{Optical magnetic imaging of living cells.}}
\newblock \emph{\bibinfo{journal}{Nature}} \textbf{\bibinfo{volume}{496}},
  \bibinfo{pages}{486--9} (\bibinfo{year}{2013}).

\bibitem{Steinert2013}
\bibinfo{author}{Steinert, S.} \emph{et~al.}
\newblock \bibinfo{title}{{Magnetic spin imaging under ambient conditions with
  sub-cellular resolution.}}
\newblock \emph{\bibinfo{journal}{Nat. Commun.}} \textbf{\bibinfo{volume}{4}},
  \bibinfo{pages}{1607} (\bibinfo{year}{2013}).

\bibitem{DeVience2015}
\bibinfo{author}{DeVience, S.~J.} \emph{et~al.}
\newblock \bibinfo{title}{{Nanoscale NMR spectroscopy and imaging of multiple
  nuclear species}}.
\newblock \emph{\bibinfo{journal}{Nat. Nanotechnol.}}
  \textbf{\bibinfo{volume}{10}}, \bibinfo{pages}{129} (\bibinfo{year}{2015}).

\bibitem{Simpson2017}
\bibinfo{author}{Simpson, D.~A.} \emph{et~al.}
\newblock \bibinfo{title}{{Electron paramagnetic resonance microscopy using
  spins in diamond under ambient conditions}}.
\newblock \emph{\bibinfo{journal}{Nat. Commun.}} \textbf{\bibinfo{volume}{8}},
  \bibinfo{pages}{458} (\bibinfo{year}{2017}).

\bibitem{Broadway2020}
\bibinfo{author}{{Broadway}, D.~A.} \emph{et~al.}
\newblock \bibinfo{title}{{Imaging domain reversal in an ultrathin van der
  Waals ferromagnet}}.
\newblock \emph{\bibinfo{journal}{pre-print}} \bibinfo{pages}{arXiv:2003.08470}
  (\bibinfo{year}{2020}).

\bibitem{Ku2019}
\bibinfo{author}{Ku, M. J.~H.} \emph{et~al.}
\newblock \bibinfo{title}{{Imaging Viscous Flow of the Dirac Fluid in Graphene
  Using a Quantum Spin Magnetometer}}  (\bibinfo{year}{2019}).
\newblock \eprint{arXiv:1905.10791v1}.

\bibitem{Tisler2013}
\bibinfo{author}{Tisler, J.} \emph{et~al.}
\newblock \bibinfo{title}{{Single defect center scanning near-field optical
  microscopy on graphene}}.
\newblock \emph{\bibinfo{journal}{Nano Letters}} \textbf{\bibinfo{volume}{13}},
  \bibinfo{pages}{3152--3156} (\bibinfo{year}{2013}).

\bibitem{Tetienne2019}
\bibinfo{author}{Tetienne, J.-P.} \emph{et~al.}
\newblock \bibinfo{title}{Apparent delocalization of the current density in
  metallic wires observed with diamond nitrogen-vacancy magnetometry}.
\newblock \emph{\bibinfo{journal}{Phys. Rev. B}} \textbf{\bibinfo{volume}{99}},
  \bibinfo{pages}{014436} (\bibinfo{year}{2019}).

\bibitem{Lillie2019}
\bibinfo{author}{Lillie, S.~E.} \emph{et~al.}
\newblock \bibinfo{title}{Imaging graphene field-effect transistors on diamond
  using nitrogen-vacancy microscopy}.
\newblock \emph{\bibinfo{journal}{Phys. Rev. Applied}}
  \textbf{\bibinfo{volume}{12}}, \bibinfo{pages}{024018}
  (\bibinfo{year}{2019}).

\bibitem{Tetienne2018b}
\bibinfo{author}{Tetienne, J.-P.} \emph{et~al.}
\newblock \bibinfo{title}{{Proximity-induced artefacts in magnetic imaging with
  nitrogen-vacancy ensembles in diamond}}.
\newblock \emph{\bibinfo{journal}{Sensors}} \textbf{\bibinfo{volume}{18}},
  \bibinfo{pages}{1290} (\bibinfo{year}{2018}).

\bibitem{Smith2019}
\bibinfo{author}{Smith, J.~M.}, \bibinfo{author}{Meynell, S.~A.},
  \bibinfo{author}{Jayich, A. C.~B.} \& \bibinfo{author}{Meijer, J.}
\newblock \bibinfo{title}{{Colour centre generation in diamond for quantum
  technologies}}.
\newblock \emph{\bibinfo{journal}{Nanophotonics}} \textbf{\bibinfo{volume}{8}},
  \bibinfo{pages}{1889--1906} (\bibinfo{year}{2019}).

\bibitem{Achard2020}
\bibinfo{author}{Achard, J.}, \bibinfo{author}{Jacques, V.} \&
  \bibinfo{author}{Tallaire, A.}
\newblock \bibinfo{title}{{Chemical vapour deposition diamond single crystals
  with nitrogen-vacancy centres: a review of material synthesis and technology
  for quantum sensing applications}}.
\newblock \emph{\bibinfo{journal}{J. Phys. D. Appl. Phys.}}
  \textbf{\bibinfo{volume}{53}}, \bibinfo{pages}{313001}
  (\bibinfo{year}{2020}).

\bibitem{Pezzagna2011a}
\bibinfo{author}{Pezzagna, S.}, \bibinfo{author}{Rogalla, D.},
  \bibinfo{author}{Wildanger, D.}, \bibinfo{author}{Meijer, J.} \&
  \bibinfo{author}{Zaitsev, A.}
\newblock \bibinfo{title}{{Creation and nature of optical centres in diamond
  for single-photon emission -- overview and critical remarks}}.
\newblock \emph{\bibinfo{journal}{New J. Phys.}} \textbf{\bibinfo{volume}{13}}
  (\bibinfo{year}{2011}).

\bibitem{Acosta2009}
\bibinfo{author}{Acosta, V.~M.} \emph{et~al.}
\newblock \bibinfo{title}{{Diamonds with a high density of nitrogen-vacancy
  centers for magnetometry applications}}.
\newblock \emph{\bibinfo{journal}{Physical Review B}}
  \textbf{\bibinfo{volume}{80}}, \bibinfo{pages}{115202}
  (\bibinfo{year}{2009}).

\bibitem{Deak2014}
\bibinfo{author}{De{\'{a}}k, P.}, \bibinfo{author}{Aradi, B.},
  \bibinfo{author}{Kaviani, M.}, \bibinfo{author}{Frauenheim, T.} \&
  \bibinfo{author}{Gali, A.}
\newblock \bibinfo{title}{{Formation of NV centers in diamond: A theoretical
  study based on calculated transitions and migration of nitrogen and vacancy
  related defects}}.
\newblock \emph{\bibinfo{journal}{Physical Review B - Condensed Matter and
  Materials Physics}} \textbf{\bibinfo{volume}{89}}, \bibinfo{pages}{075203}
  (\bibinfo{year}{2014}).

\bibitem{Tetienne2018a}
\bibinfo{author}{Tetienne, J.-P.} \emph{et~al.}
\newblock \bibinfo{title}{{Spin properties of dense near-surface ensembles of
  nitrogen-vacancy centers in diamond}}.
\newblock \emph{\bibinfo{journal}{Physical Review B}}
  \textbf{\bibinfo{volume}{97}}, \bibinfo{pages}{085402}
  (\bibinfo{year}{2018}).

\bibitem{Barry2019}
\bibinfo{author}{{Barry}, J.~F.} \emph{et~al.}
\newblock \bibinfo{title}{{Sensitivity Optimization for NV-Diamond
  Magnetometry}}.
\newblock \emph{\bibinfo{journal}{Reviews of Modern Physics}}
  \textbf{\bibinfo{volume}{92}}, \bibinfo{pages}{15004} (\bibinfo{year}{2020}).

\bibitem{Dreau2011}
\bibinfo{author}{Dr{\'{e}}au, a.} \emph{et~al.}
\newblock \bibinfo{title}{{Avoiding power broadening in optically detected
  magnetic resonance of single NV defects for enhanced dc magnetic field
  sensitivity}}.
\newblock \emph{\bibinfo{journal}{Physical Review B - Condensed Matter and
  Materials Physics}} \textbf{\bibinfo{volume}{84}}, \bibinfo{pages}{195204}
  (\bibinfo{year}{2011}).

\bibitem{Taylor2008}
\bibinfo{author}{Taylor, J.~M.} \emph{et~al.}
\newblock \bibinfo{title}{{High-sensitivity diamond magnetometer with nanoscale
  resolution}}.
\newblock \emph{\bibinfo{journal}{Nat. Phys.}} \textbf{\bibinfo{volume}{4}},
  \bibinfo{pages}{810--816} (\bibinfo{year}{2008}).

\bibitem{Huang2013}
\bibinfo{author}{Huang, Z.} \emph{et~al.}
\newblock \bibinfo{title}{{Diamond nitrogen-vacancy centers created by scanning
  focused helium ion beam and annealing}}.
\newblock \emph{\bibinfo{journal}{Applied Physics Letters}}
  \textbf{\bibinfo{volume}{103}}, \bibinfo{pages}{081906}
  (\bibinfo{year}{2013}).

\bibitem{McCloskey2014}
\bibinfo{author}{McCloskey, D.} \emph{et~al.}
\newblock \bibinfo{title}{{Helium ion microscope generated nitrogen-vacancy
  centres in type Ib diamond}}.
\newblock \emph{\bibinfo{journal}{Applied Physics Letters}}
  \textbf{\bibinfo{volume}{104}}, \bibinfo{pages}{031109}
  (\bibinfo{year}{2014}).

\bibitem{Fescenko2019}
\bibinfo{author}{Fescenko, I.} \emph{et~al.}
\newblock \bibinfo{title}{{Diamond Magnetic Microscopy of Malarial Hemozoin
  Nanocrystals}}.
\newblock \emph{\bibinfo{journal}{Phys. Rev. Appl.}}
  \textbf{\bibinfo{volume}{11}}, \bibinfo{pages}{034029}
  (\bibinfo{year}{2019}).

\bibitem{Ohno2012}
\bibinfo{author}{Ohno, K.} \emph{et~al.}
\newblock \bibinfo{title}{Engineering shallow spins in diamond with nitrogen
  delta-doping}.
\newblock \emph{\bibinfo{journal}{Applied Physics Letters}}
  \textbf{\bibinfo{volume}{101}}, \bibinfo{pages}{082413}
  (\bibinfo{year}{2012}).

\bibitem{Ohno2014}
\bibinfo{author}{Ohno, K.} \emph{et~al.}
\newblock \bibinfo{title}{{Three-dimensional localization of spins in diamond
  using 12C implantation}}.
\newblock \emph{\bibinfo{journal}{Applied Physics Letters}}
  \textbf{\bibinfo{volume}{105}}, \bibinfo{pages}{052406}
  (\bibinfo{year}{2014}).

\bibitem{FavarodeOliveira2016}
\bibinfo{author}{{F{\'{a}}varo de Oliveira}, F.} \emph{et~al.}
\newblock \bibinfo{title}{{Toward Optimized Surface $\delta$-Profiles of
  Nitrogen-Vacancy Centers Activated by Helium Irradiation in Diamond}}.
\newblock \emph{\bibinfo{journal}{Nano Letters}} \textbf{\bibinfo{volume}{16}},
  \bibinfo{pages}{2228--2233} (\bibinfo{year}{2016}).

\bibitem{Kleinsasser2016}
\bibinfo{author}{Kleinsasser, E.~E.} \emph{et~al.}
\newblock \bibinfo{title}{{High density NV sensing surface created via
  He{\^{}}(+) ion implantation of (12){\^{}}C diamond}}.
\newblock \emph{\bibinfo{journal}{Appl. Phys. Lett.}}
  \textbf{\bibinfo{volume}{108}}, \bibinfo{pages}{202401}
  (\bibinfo{year}{2016}).

\bibitem{Kim2012}
\bibinfo{author}{Kim, E.}, \bibinfo{author}{Acosta, V.~M.},
  \bibinfo{author}{Bauch, E.}, \bibinfo{author}{Budker, D.} \&
  \bibinfo{author}{Hemmer, P.~R.}
\newblock \bibinfo{title}{Electron spin resonance shift and linewidth
  broadening of nitrogen-vacancy centers in diamond as a function of electron
  irradiation dose}.
\newblock \emph{\bibinfo{journal}{Applied Physics Letters}}
  \textbf{\bibinfo{volume}{101}}, \bibinfo{pages}{082410}
  (\bibinfo{year}{2012}).

\bibitem{McLellan2016}
\bibinfo{author}{McLellan, C.~A.} \emph{et~al.}
\newblock \bibinfo{title}{{Patterned Formation of Highly Coherent
  Nitrogen-Vacancy Centers Using a Focused Electron Irradiation Technique}}.
\newblock \emph{\bibinfo{journal}{Nano Letters}} \textbf{\bibinfo{volume}{16}},
  \bibinfo{pages}{2450--2454} (\bibinfo{year}{2016}).

\bibitem{Pezzagna2010}
\bibinfo{author}{Pezzagna, S.}, \bibinfo{author}{Naydenov, B.},
  \bibinfo{author}{Jelezko, F.}, \bibinfo{author}{Wrachtrup, J.} \&
  \bibinfo{author}{Meijer, J.}
\newblock \bibinfo{title}{{Creation efficiency of nitrogen-vacancy centres in
  diamond}}.
\newblock \emph{\bibinfo{journal}{New Journal of Physics}}
  \textbf{\bibinfo{volume}{12}}, \bibinfo{pages}{065017}
  (\bibinfo{year}{2010}).

\bibitem{Toyli2010}
\bibinfo{author}{Toyli, D.~M.}, \bibinfo{author}{Weis, C.~D.},
  \bibinfo{author}{Fuchs, G.~D.}, \bibinfo{author}{Schenkel, T.} \&
  \bibinfo{author}{Awschalom, D.~D.}
\newblock \bibinfo{title}{{Chip-scale nanofabrication of single spins and spin
  arrays in diamond}}.
\newblock \emph{\bibinfo{journal}{Nano Letters}} \textbf{\bibinfo{volume}{10}},
  \bibinfo{pages}{3168--3172} (\bibinfo{year}{2010}).

\bibitem{Spinicelli2011}
\bibinfo{author}{Spinicelli, P.} \emph{et~al.}
\newblock \bibinfo{title}{{Engineered arrays of nitrogen-vacancy color centers
  in diamond based on implantation of CN$^-$ Molecules through nanoapertures}}.
\newblock \emph{\bibinfo{journal}{New Journal of Physics}}
  \textbf{\bibinfo{volume}{13}}, \bibinfo{pages}{025014}
  (\bibinfo{year}{2011}).

\bibitem{Antonov2014}
\bibinfo{author}{Antonov, D.} \emph{et~al.}
\newblock \bibinfo{title}{{Statistical investigations on nitrogen-vacancy
  center creation}}.
\newblock \emph{\bibinfo{journal}{Applied Physics Letters}}
  \textbf{\bibinfo{volume}{104}}, \bibinfo{pages}{012105}
  (\bibinfo{year}{2014}).

\bibitem{DeOliveira2017}
\bibinfo{author}{de~Oliveira, F.~F.} \emph{et~al.}
\newblock \bibinfo{title}{{Tailoring spin defects in diamond}}.
\newblock \emph{\bibinfo{journal}{Nat. Comm.}} \textbf{\bibinfo{volume}{8}},
  \bibinfo{pages}{15409} (\bibinfo{year}{2017}).

\bibitem{Bauch2019}
\bibinfo{author}{{Bauch}, E.} \emph{et~al.}
\newblock \bibinfo{title}{{Decoherence of dipolar spin ensembles in diamond}}.
\newblock \emph{\bibinfo{journal}{pre-print}} \bibinfo{pages}{arXiv:1904.08763}
  (\bibinfo{year}{2019}).
\newblock \eprint{1904.08763}.

\bibitem{Tallaire2004}
\bibinfo{author}{Tallaire, A.} \emph{et~al.}
\newblock \bibinfo{title}{{Oxygen plasma pre-treatments for high quality
  homoepitaxial CVD diamond deposition}}.
\newblock \emph{\bibinfo{journal}{Phys. Status Solidi}}
  \textbf{\bibinfo{volume}{201}}, \bibinfo{pages}{2419--2424}
  (\bibinfo{year}{2004}).

\bibitem{Edmonds2012}
\bibinfo{author}{Edmonds, A.~M.} \emph{et~al.}
\newblock \bibinfo{title}{{Production of oriented nitrogen-vacancy color
  centers in synthetic diamond}}.
\newblock \emph{\bibinfo{journal}{Phys. Rev. B - Condens. Matter Mater. Phys.}}
  \textbf{\bibinfo{volume}{86}}, \bibinfo{pages}{035201}
  (\bibinfo{year}{2012}).

\bibitem{Teraji2015}
\bibinfo{author}{Teraji, T.}
\newblock \bibinfo{title}{{High-quality and high-purity homoepitaxial diamond (
  100 ) film growth under high oxygen concentration condition}}.
\newblock \emph{\bibinfo{journal}{J. Appl. Phys.}}
  \textbf{\bibinfo{volume}{118}}, \bibinfo{pages}{115304}
  (\bibinfo{year}{2015}).

\bibitem{McCoey2020}
\bibinfo{author}{McCoey, J.~M.} \emph{et~al.}
\newblock \bibinfo{title}{{Quantum Magnetic Imaging of Iron Biomineralization
  in Teeth of the Chiton Acanthopleura hirtosa}}.
\newblock \emph{\bibinfo{journal}{Small Methods}} \textbf{\bibinfo{volume}{4}},
  \bibinfo{pages}{1900754} (\bibinfo{year}{2020}).

\bibitem{Lillie2020}
\bibinfo{author}{Lillie, S.~E.} \emph{et~al.}
\newblock \bibinfo{title}{{Laser Modulation of Superconductivity in a Cryogenic
  Wide- fi eld Nitrogen-Vacancy Microscope}}.
\newblock \emph{\bibinfo{journal}{Nano Lett.}} \textbf{\bibinfo{volume}{20}},
  \bibinfo{pages}{1855--1861} (\bibinfo{year}{2020}).

\bibitem{Bluvstein2019}
\bibinfo{author}{Bluvstein, D.}, \bibinfo{author}{Zhang, Z.} \&
  \bibinfo{author}{Jayich, A. C.~B.}
\newblock \bibinfo{title}{{Identifying and Mitigating Charge Instabilities in
  Shallow Diamond Nitrogen-Vacancy Centers}}.
\newblock \emph{\bibinfo{journal}{Phys. Rev. Lett.}}
  \textbf{\bibinfo{volume}{122}}, \bibinfo{pages}{76101}
  (\bibinfo{year}{2019}).

\bibitem{Manson2018}
\bibinfo{author}{Manson, N.~B.}, \bibinfo{author}{Hedges, M.},
  \bibinfo{author}{Barson, M. S.~J.}, \bibinfo{author}{Ahlefeldt, R.} \&
  \bibinfo{author}{Doherty, M.~W.}
\newblock \bibinfo{title}{{NV$^-$–N$^+$ pair centre in 1b diamond}}.
\newblock \emph{\bibinfo{journal}{New J. Phys.}} \textbf{\bibinfo{volume}{20}},
  \bibinfo{pages}{113037} (\bibinfo{year}{2018}).

\bibitem{Stacey2019}
\bibinfo{author}{Stacey, A.} \emph{et~al.}
\newblock \bibinfo{title}{{Evidence for Primal sp 2 Defects at the Diamond
  Surface : Candidates for Electron Trapping and Noise Sources}}.
\newblock \emph{\bibinfo{journal}{Adv. Mater. Interfaces}}
  \textbf{\bibinfo{volume}{6}}, \bibinfo{pages}{1801449}
  (\bibinfo{year}{2019}).

\bibitem{Naydenov2011}
\bibinfo{author}{Naydenov, B.} \emph{et~al.}
\newblock \bibinfo{title}{{Dynamical Decoupling of a single-electron spin at
  room temperature}}.
\newblock \emph{\bibinfo{journal}{Phys. Rev. B}} \textbf{\bibinfo{volume}{83}},
  \bibinfo{pages}{081201} (\bibinfo{year}{2011}).

\bibitem{DeLange2011}
\bibinfo{author}{de~Lange, G.}, \bibinfo{author}{Riste, D.},
  \bibinfo{author}{Dobrovitski, V.~V.} \& \bibinfo{author}{Hanson, R.}
\newblock \bibinfo{title}{{Single-Spin Magnetometry with Multipulse Sensing
  Sequences}}.
\newblock \emph{\bibinfo{journal}{Phys. Rev. Lett.}}
  \textbf{\bibinfo{volume}{106}}, \bibinfo{pages}{080802}
  (\bibinfo{year}{2011}).

\bibitem{DeLange2010}
\bibinfo{author}{de~Lange, G.}, \bibinfo{author}{Wang, Z.~H.},
  \bibinfo{author}{Rist{\`{e}}, D.}, \bibinfo{author}{Dobrovitski, V.~V.} \&
  \bibinfo{author}{Hanson, R.}
\newblock \bibinfo{title}{{Universal dynamical decoupling of a single
  solid-state spin from a spin bath.}}
\newblock \emph{\bibinfo{journal}{Science}} \textbf{\bibinfo{volume}{330}},
  \bibinfo{pages}{60--63} (\bibinfo{year}{2010}).

\bibitem{Bar-Gill2013}
\bibinfo{author}{Bar-Gill, N.}, \bibinfo{author}{Pham, L.~M.},
  \bibinfo{author}{Jarmola, A.}, \bibinfo{author}{Budker, D.} \&
  \bibinfo{author}{Walsworth, R.~L.}
\newblock \bibinfo{title}{{Solid-state electronic spin coherence time
  approaching one second}}.
\newblock \emph{\bibinfo{journal}{Nat. Commun.}} \textbf{\bibinfo{volume}{4}},
  \bibinfo{pages}{1743} (\bibinfo{year}{2013}).

\bibitem{Romach2015}
\bibinfo{author}{Romach, Y.} \emph{et~al.}
\newblock \bibinfo{title}{{Spectroscopy of surface-induced noise using shallow
  spins in diamond}}.
\newblock \emph{\bibinfo{journal}{Phys. Rev. Lett.}}
  \textbf{\bibinfo{volume}{114}}, \bibinfo{pages}{017601}
  (\bibinfo{year}{2015}).

\bibitem{Broadway2019}
\bibinfo{author}{Broadway, D.~A.} \emph{et~al.}
\newblock \bibinfo{title}{{Microscopic Imaging of the Stress Tensor in Diamond
  Using in Situ Quantum Sensors}}.
\newblock \emph{\bibinfo{journal}{Nano Lett.}} \textbf{\bibinfo{volume}{19}},
  \bibinfo{pages}{4543--4550} (\bibinfo{year}{2019}).

\bibitem{Kehayias2019}
\bibinfo{author}{Kehayias, P.} \emph{et~al.}
\newblock \bibinfo{title}{{Imaging crystal stress in diamond using ensembles of
  nitrogen-vacancy centers}}.
\newblock \emph{\bibinfo{journal}{Phys. Rev. B - Condens. Matter Mater. Phys.}}
  \textbf{\bibinfo{volume}{100}}, \bibinfo{pages}{174103}
  (\bibinfo{year}{2019}).

\bibitem{Friel2009}
\bibinfo{author}{Friel, I.} \emph{et~al.}
\newblock \bibinfo{title}{{Diamond {\&} Related Materials Control of surface
  and bulk crystalline quality in single crystal diamond grown by chemical
  vapour deposition}}.
\newblock \emph{\bibinfo{journal}{Diam. Relat. Mater.}}
  \textbf{\bibinfo{volume}{18}}, \bibinfo{pages}{808--815}
  (\bibinfo{year}{2009}).

\bibitem{Appel2016}
\bibinfo{author}{Appel, P.} \emph{et~al.}
\newblock \bibinfo{title}{{Fabrication of all diamond scanning probes for
  nanoscale magnetometry}}.
\newblock \emph{\bibinfo{journal}{Rev. Sci. Instrum.}}
  \textbf{\bibinfo{volume}{87}}, \bibinfo{pages}{063703}
  (\bibinfo{year}{2016}).

\bibitem{Sangtawesin2019}
\bibinfo{author}{Sangtawesin, S.} \emph{et~al.}
\newblock \bibinfo{title}{{Origins of Diamond Surface Noise Probed by
  Correlating Single-Spin Measurements with Surface Spectroscopy}}.
\newblock \emph{\bibinfo{journal}{Phys. Rev. X}} \textbf{\bibinfo{volume}{9}},
  \bibinfo{pages}{031052} (\bibinfo{year}{2019}).

\bibitem{Martineau2009}
\bibinfo{author}{Martineau, P.~M.} \emph{et~al.}
\newblock \bibinfo{title}{{High crystalline quality single crystal chemical
  vapour deposition diamond}}.
\newblock \emph{\bibinfo{journal}{J. Phys. Condens. Matter}}
  \textbf{\bibinfo{volume}{21}}, \bibinfo{pages}{364205}
  (\bibinfo{year}{2009}).

\bibitem{privcom}
\bibinfo{author}{Maletinsky, P.} \emph{et~al.}
\newblock \emph{\bibinfo{journal}{Private communication}} .

\bibitem{Tetienne2012}
\bibinfo{author}{Tetienne, J.~P.} \emph{et~al.}
\newblock \bibinfo{title}{{Magnetic-field-dependent photodynamics of single NV
  defects in diamond: An application to qualitative all-optical magnetic
  imaging}}.
\newblock \emph{\bibinfo{journal}{New J. Phys.}} \textbf{\bibinfo{volume}{14}},
  \bibinfo{pages}{103033} (\bibinfo{year}{2012}).

\bibitem{Stacey2012a}
\bibinfo{author}{Stacey, A.} \emph{et~al.}
\newblock \bibinfo{title}{{Depletion of nitrogen-vacancy color centers in
  diamond via hydrogen passivation}}.
\newblock \emph{\bibinfo{journal}{Appl. Phys. Lett.}}
  \textbf{\bibinfo{volume}{100}}, \bibinfo{pages}{071902}
  (\bibinfo{year}{2012}).

\bibitem{Kucsko2018}
\bibinfo{author}{Kucsko, G.} \emph{et~al.}
\newblock \bibinfo{title}{{Critical Thermalization of a Disordered Dipolar Spin
  System in Diamond}}.
\newblock \emph{\bibinfo{journal}{Phys. Rev. Lett.}}
  \textbf{\bibinfo{volume}{121}}, \bibinfo{pages}{23601}
  (\bibinfo{year}{2018}).

\bibitem{DeOliveira2015}
\bibinfo{author}{de~Oliveira, F.~F.} \emph{et~al.}
\newblock \bibinfo{title}{{Effect of Low-Damage Inductively Coupled Plasma on
  Shallow NV Centers in Diamond}}.
\newblock \emph{\bibinfo{journal}{Appl. Phys. Lett.}}
  \textbf{\bibinfo{volume}{107}}, \bibinfo{pages}{073107}
  (\bibinfo{year}{2015}).

\bibitem{Luhmann}
\bibinfo{author}{L{\"{u}}hmann, T.}, \bibinfo{author}{John, R.},
  \bibinfo{author}{Wunderlich, R.}, \bibinfo{author}{Meijer, J.} \&
  \bibinfo{author}{Pezzagna, S.}
\newblock \bibinfo{title}{{Coulomb-driven single defect engineering for
  scalable qubits and spin sensors in diamond}}.
\newblock \emph{\bibinfo{journal}{Nat. Commun.}} \textbf{\bibinfo{volume}{10}},
  \bibinfo{pages}{4956} (\bibinfo{year}{2019}).

\bibitem{Herbschleb}
\bibinfo{author}{Herbschleb, E.~D.} \emph{et~al.}
\newblock \bibinfo{title}{{Ultra-long coherence times amongst room-temperature
  solid-state spins}}.
\newblock \emph{\bibinfo{journal}{Nat. Commun.}} \textbf{\bibinfo{volume}{10}},
  \bibinfo{pages}{3766} (\bibinfo{year}{2019}).

\end{thebibliography}

\clearpage

\end{document}